\documentclass[useAMS,usenatbib]{mn2e}
\usepackage{graphicx,natbib,color,multirow,amsmath,deluxetable}
\definecolor{titlecol}{rgb}{0,0,1}
\definecolor{titlecol2}{rgb}{0,0.65,0}
\definecolor{titlecol3}{rgb}{0.99,0.4,0.}

\def\builderUC       {} % builder, unconfirmed

\def\ffeat            {$f_{\rm features}$}
\def\mffeat           {f_{\rm features}}
\def\fsmooth      {$f_{\rm smooth}$}
\def\mfsmooth    {f_{\rm smooth}}

\def\kappamean {$\overline{\kappa}$}
\def\mkappamean {\overline{\kappa}}
\def\stellarity   {{\tt CLASS\_STAR}}
\def\musb       {$\mu_{\rm{\textsc{sb}}}$}
\def\mmusb     {\mu_{\rm{\textsc{sb}}}}

\def\lesssim{\mathrel{\hbox{\rlap{\hbox{\lower3pt\hbox{$\sim$}}}\hbox{\raise2pt\hbox{$<$}}}}}
\def\gtrsim{\mathrel{\hbox{\rlap{\hbox{\lower3pt\hbox{$\sim$}}}\hbox{\raise2pt\hbox{$>$}}}}}

\makeatletter
\def\@xfootnote[#1]{%
  \protected@xdef\@thefnmark{#1}%
  \@footnotemark\@footnotetext}
\makeatother

\begin{document}

\title[Galaxy Zoo CANDELS Data Release]{Galaxy Zoo: Quantitative Visual Morphological Classifications for 48,000 galaxies from CANDELS\thanks{This publication has been made possible by the participation of more than 95,000 volunteers in the Galaxy Zoo project. The contributions of the more than $40,000$ of those who registered a username with Galaxy Zoo are
individually acknowledged at http://authors.galaxyzoo.org/ .} } 

\author[Simmons et al.]{\parbox[t]{16cm}{B. D. Simmons$^{1,2,3}$\thanks{E-mail: bdsimmons@ucsd.edu}\footnote{Einstein Fellow}, 
Chris Lintott$^{1}$, 
Kyle W. Willett$^{4,5}$, 
Karen L. Masters$^{6,7}$, 
Jeyhan S. Kartaltepe$^{8}$, 
Boris H\"au\ss ler$^{1,9,10}$,
Sugata Kaviraj$^{9, 11}$,
Coleman Krawczyk$^{6}$,
S. J. Kruk$^{1}$,
Daniel H. McIntosh$^{12}$,
R. J. Smethurst$^{1}$, 
Robert C. Nichol$^{6,7}$, 
Claudia Scarlata$^{4}$,
Kevin Schawinski$^{13}$, 
Christopher J. Conselice$^{14}$,
Omar Almaini$^{14}$, 
Henry C. Ferguson$^{15}$,
Lucy Fortson$^{4}$, 
William Hartley$^{13}$, 
Dale Kocevski$^{16}$,
Anton M. Koekemoer$^{15}$,
Alice Mortlock$^{14}$, 
Jeffrey A. Newman$^{17}$,
Steven P. Bamford$^{14}$,
N. A. Grogin$^{15}$,
Ray A. Lucas$^{15}$,
Nimish P. Hathi$^{18}$,
Elizabeth McGrath$^{16}$,
Michael Peth$^{19}$,
Janine Pforr $^{18, 20}$,
Zachary Rizer$^{12}$,
Stijn Wuyts$^{21}$,
Guillermo Barro$^{25}$,
{\builderUC Eric F. Bell$^{22}$,}
Marco Castellano$^{23}$,
{\builderUC Tomas Dahlen$^{15}$,} 
Avishai Dekel$^{24}$
{\builderUC Jamie Ownsworth$^{14}$,}
{\builderUC Sandra M. Faber$^{25}$,} %UCO/Lick
{\builderUC Steven L. Finkelstein$^{26}$,} 
{\builderUC Adriano Fontana$^{23}$,}
{\builderUC Audrey Galametz$^{27}$,}
{\builderUC Ruth Gr\"utzbauch$^{14, 28}$,} 
{\builderUC David Koo$^{25}$,} %UCSC
{\builderUC Jennifer Lotz$^{15}$,} %ST
{\builderUC Bahram Mobasher$^{29}$,} %UCR / check
{\builderUC Mark Mozena$^{25}$,}
{\builderUC Mara Salvato$^{27}$,} 
{\builderUC Tommy Wiklind$^{30}$} 
\vspace{0.1in} }\\
$^{1}$Oxford Astrophysics, Denys Wilkinson Building, Keble Road, Oxford OX1 3RH, UK\\
$^{2}$Balliol College, Oxford, UK\\
$^{3}$Center for Astrophysics and Space Sciences (CASS), Department of Physics, University of California, San Diego, CA 92093, USA\\
$^{4}$School of Physics and Astronomy, University of Minnesota, 116 Church St. SE, Minneapolis, MN 55455, USA\\
$^{5}$Department of Physics and Astronomy, University of Kentucky, Lexington, KY 40506, USA\\
$^{6}$Institute of Cosmology \& Gravitation, University of Portsmouth, Dennis Sciama Building, Portsmouth PO1 3FX, UK\\
$^{7}$SEPnet,\thanks{www.sepnet.ac.uk} South East Physics Network\\
$^{8}$School of Physics and Astronomy, Rochester Institute of Technology, 84 Lomb Memorial Drive, Rochester, NY 14623, USA\\
$^{9}$Centre for Astrophysics Research, University of Hertfordshire, College Lane, Hatfield AL10 9AB, UK\\
$^{10}$European Southern Observatory, Alonso de Cordova 3107, Vitacura, Casilla 19001, Santiago, Chile\\
$^{11}$Worcester College, Oxford, UK\\
$^{12}$Department of Physics \& Astronomy, University of Missouri-Kansas City, 5110 Rockhill Rd., Kansas City, MO 64110, USA\\
$^{13}$Institute for Astronomy, ETH Z\"urich, Wolfgang-Pauli-Strasse 27, CH-8093 Z\"urich, Switzerland\\
$^{14}$School of Physics \& Astronomy, The University of Nottingham, University Park, Nottingham, NG7 2RD, UK\\
$^{15}$Space Telescope Science Institute, 3700 San Martin Drive, Baltimore, MD 21218\\
$^{16}$Department of Physics and Astronomy, Colby College, Waterville, ME 04901, USA\\
$^{17}$Department of Physics and Astronomy \& PITT PACC, University of Pittsburgh, Pittsburgh, PA 15217, USA\\
$^{18}$Aix Marseille Universit\'e, CNRS, LAM (Laboratoire d'Astrophysique de Marseille) UMR 7326, F-13388 Marseille, France\\
$^{19}$Department of Physics and Astronomy, The Johns Hopkins University, Baltimore, MD 21218, USA\\
$^{20}$Scientific Support Office, Directorate of Science and Robotic Exploration, European Space Research and Technology Centre (ESA/ESTEC), Keplerlaan 1, 2201 AZ Noordwijk, The Netherlands\\
$^{21}$Department of Physics, University of Bath, Claverton Down, Bath BA2 7AY, UK \\
$^{22}$Department of Astronomy, University of Michigan, Ann Arbor, MI 48104, USA\\
$^{23}$INAF-Osservatorio Astronomico di Roma, Via Frascati 33, I-00040, Monte Porzio Catone, Italy\\
$^{24}$Center for Astrophysics and Planetary Science, Racah Institute of Physics, The Hebrew University, Jerusalem 91904, Israel\\
$^{25}$UC Observatories/Lick Observatory and Department of Astronomy and Astrophysics, University of California, Santa Cruz, CA 95064, USA\\
$^{26}$Department of Astronomy, The University of Texas at Austin, Austin, TX 78712, USA\\
$^{27}$Max-Planck-Institut f{\"u}r extraterrestrische Physik, Giessenbachstrasse 1, DÐ85748 Garching bei M{\"u}nchen, Germany\\
$^{28}$Centre for Astronomy and Astrophysics, University of Lisbon, P-1349-018 Lisbon, Portugal\\
$^{29}$Department of Physics \& Astronomy, University of California, Riverside, CA 92521, USA\\
$^{30}$Joint ALMA Observatory, Alonso de Cordova 3107, Vitacura, Santiago, Chile
   }

\maketitle
  
\label{firstpage}
  
\clearpage

\begin{abstract}

We present quantified visual morphologies of approximately 48,000 galaxies observed in three \emph{Hubble Space Telescope} legacy fields by the Cosmic And Near-infrared Deep Extragalactic Legacy Survey (CANDELS) and classified by participants in the Galaxy Zoo project. 90\% of galaxies have $z \leq 3$ and are observed in rest-frame optical wavelengths by CANDELS. Each galaxy received an average of 40 independent classifications, which we combine into detailed morphological information on galaxy features such as clumpiness, bar instabilities, spiral structure, and merger and tidal signatures. We apply a consensus-based classifier weighting method that preserves classifier independence while effectively down-weighting significantly outlying classifications. After analysing the effect of varying image depth on reported classifications, we also provide depth-corrected classifications which both preserve the information in the deepest observations and also enable the use of classifications at comparable depths across the full survey. Comparing the Galaxy Zoo classifications to previous classifications of the same galaxies shows very good agreement; for some applications the high number of independent classifications provided by Galaxy Zoo provides an advantage in selecting galaxies with a particular morphological profile, while in others the combination of Galaxy Zoo with other classifications is a more promising approach than using any one method alone. We combine the Galaxy Zoo classifications of ``smooth'' galaxies with parametric morphologies to select a sample of featureless disks at $1 \leq z \leq 3$, which may represent a dynamically warmer progenitor population to the settled disk galaxies seen at later epochs.

  \end{abstract}
  

%%%%%%%%%%%%%%%%%%%%%%%%%%%%%%%%%%%%%%%%%%%%%%
%
%  
\section{Introduction}
%
%
%%%%%%%%%%%%%%%%%%%%%%%%%%%%%%%%%%%%%%%%%%%%%%

The shape and appearance of a galaxy reflect the underlying physical processes that have formed it and which continue to influence its evolution. For example, the signatures of past merger events \citep[from $z \sim 2$ onwards;][]{martig12} are thought to be visible even at $z = 0$ in the form of a galactic bulge; the strength of the bulge is thought to be tied to the strength of the merger, as indeed the lack of a bulge indicates a lack of significant mergers \citep[e.g.,][]{kormendy10}. Likewise, other morphological features are tied to disk instabilities and resonances \citep[e.g.,][]{kormendy04,b_elmegreen08,donghia13}, and orbital changes from the disruptive \citep[mergers; e.g.,][]{darg10b,darg10a,lotz08a,lotz08b} to the relatively subtle \citep[e.g., bars,][and for studies of visually-identified bars at $z > 0$ specifically, see e.g. \citeauthor{sheth08} \citeyear{sheth08}; \citeauthor{melvin14} \citeyear{melvin14}; \citeauthor{simmons14} \citeyear{simmons14}; \citeauthor{cheung15} \citeyear{cheung15}]{athanassoula92b,sellwood93,athanassoula05,athanassoula13}. Combinations of morphological parameters with other measures, such as environment, color, mass and star formation histories \citep[e.g.][]{bamford09,tojeiro07,schawinski14,kaviraj14a,kaviraj14b,smethurst15} can provide more insight than either alone.

Morphological measures have a long history in astronomy \citep[e.g.,][]{hubble26,devaucouleurs,devaucouleurs59,sandage61,vandenbergh76,abraham96c,nair10}. The computerized era of astrophysics has brought with it a number of automated morphological classification techniques. Some use multiple parameters to characterise a galaxy's distribution of light \citep{sersic68,odewahn02}, while others adopt a non-parametric approach, each reducing a galaxy to one number \citep[and often used in combination; e.g.][]{abraham94, conselice03, lotz04a}. Both types of analyses lend themselves relatively well to large-scale processing of images from galaxy surveys \citep[e.g.][]{simard02,scarlata07,simard09,simard11,griffith12,lackner12,lackner13,meert15,meert16} and provide a uniform quantitative set of measures. Modern machine learning techniques, with appropriate training, are also applicable to large data sets \citep{huertascompany08,huertascompany15,dieleman15}.

However, no computer has yet exceeded the human brain's capacity for pattern detection and serendipitous discovery. Visual morphologies remain among the most nuanced and powerful measures of galaxy structure. Galaxy Zoo combines the strengths of visual classification with the volume of computer-driven approaches, using the World Wide Web to collect more independent and complete visual classifications than any group of professional astronomers is realistically capable of and combining these classifications via tested and proven techniques.

Since 2007, Galaxy Zoo has been a unique resource of quantitative and statistically robust visual galaxy morphologies. Prior to Galaxy Zoo CANDELS, three Galaxy Zoo projects have collected morphologies for over $1,000,000$ galaxies using the largest surveys to date to $z \sim 1$ \citep[][K. Willett et al., in preparation]{lintott08,lintott11,willett13}. These projects have been and continue to be extremely scientifically productive, both for the project team \citep[e.g.,][]{keel15,galloway15,willett15} and for the larger scientific community \citep[e.g.][]{amorin10,finkelman12,robaina12,combes13,joachimi15,zhang15,lopezcorredoira16}.

This paper presents morphological classifications of $49,555$ images from the Cosmic And Near-infrared Deep Extragalactic Legacy Survey \citep[CANDELS;][]{grogin11,koekemoer11}; the largest near-infrared \emph{Hubble Space Telescope (HST)} survey to date, which images galaxies at rest-frame optical wavelengths to $z \approx 3$. The morphologies are quantified by the Galaxy Zoo{\footnote{zoo4.galaxyzoo.org}} project \citep{lintott08}. Over 95,000 volunteers have contributed over 2,000,000 detailed galaxy classifications to this effort. We combine, on average, 43 independent classifications of each galaxy to produce detailed, quantitative morphological descriptions of these distant galaxies along many physical axes of interest. 

In Section \ref{sec:data} we describe the observational data and the preparation of CANDELS images for use in Galaxy Zoo. In Section \ref{sec:classifications} we detail the collection of morphological classifications and the method of weighting and combining independent classifications for each galaxy. Section \ref{sec:comparison} compares Galaxy Zoo classifications to other morphological measurements. In Section \ref{sec:result} we show an example result using the classifications, and in Section \ref{sec:summary} we summarize. Throughout
this paper we use the AB magnitude system, and where necessary we adopt a cosmology consistent with $\Lambda$CDM, with $H_{\rm 0}=70~{\rm
km~s^{-1}}$Mpc$^{\rm -1}$, $\Omega_{\rm m}=0.3$ and $\Omega_{\rm \Lambda}=0.7$ \citep{bennett13}.

%%%%%%%%%%%%%%%%%%%%%%%%%%%%%%%%%%%%%%%%%%%%%%
%
%
\section{Observational Data}\label{sec:data}
%
%
%%%%%%%%%%%%%%%%%%%%%%%%%%%%%%%%%%%%%%%%%%%%%%

\subsection{Images}\label{sec:images}

The Cosmic Assembly Near-infrared Extragalactic Legacy Survey \citep{grogin11,koekemoer11} is an \emph{HST} Treasury programme combining optical and near-infrared imaging from the Advanced Camera for Surveys (ACS) and Wide Field Camera 3 (infrared channel; WFC3/IR), providing an unprecedented opportunity to study galaxy structure and evolution across a range of redshifts. CANDELS covers the area included in five fields which had been targeted for previous studies {(GOODS-North and -South, \citeauthor{giavalisco04} \citeyear{giavalisco04}; EGS, \citeauthor{davis07} \citeyear{davis07}; UDS, \citeauthor{lawrence07} \citeyear{lawrence07}, \citeauthor{cirasuolo07} \citeyear{cirasuolo07}; and COSMOS, \citeauthor{scoville07} \citeyear{scoville07}), divided into `deep' and `wide' fields. Each of the wide fields (UDS, COSMOS, EGS and flanking fields to the GOODS-S and GOODS-N deep fields) are imaged over 2 orbits in WFC3/IR, split in a 2:1 ratio between filters F160W and F125W respectively, with parallel exposures made in F606W and F814W using ACS. Each of the deep fields (corresponding to those targeted by GOODS-S and GOODS-N) are imaged over at least 4 orbits each in both the F160W and F125W filters and 3 orbits in the F105W filter, with ACS exposures in F606W and F814W in parallel. These data are reduced and combined to produce a single mosaic for each field in each band, with drizzled resolutions of $0.03^{\prime\prime}$ and $0.06^{\prime\prime}$ per pixel for ACS and WFC3/IR, respectively \citep{koekemoer11}.

The 4th phase of Galaxy Zoo included all detections with $H \leq 25.5$ from COSMOS, GOODS-South and UDS, comprising 49,555 unique images. These were shown to visitors to the website {\tt www.galaxyzoo.org} starting on 10~September~2012. The images shown on the site were colour composites of ACS $I$ ($F814W$), WFC3 $J$ ($F125W$), and WFC3 $H$ ($F160W$) filters for the blue, green and red channels, respectively. Previous iterations of Galaxy Zoo \citep{lintott08} showed that the effect of using colour images (rather than monochrome or single filter images) for classifications is small, but that their inclusion greatly increases classifier engagement, resulting in significantly faster collection of quantitative visual morphologies. 

The physical angular sizes of the Galaxy Zoo CANDELS images were matched in different filters, using the native point-spread functions (PSFs). The images were combined with an asinh stretch \citep[described in detail in][]{lupton04} with a non-linearity value of 4.0, chosen to show clear features across a wide dynamic range. 

Sources in the dataset vary greatly in size and surface brightness, and a single set of values for channel scalings is not adequate to capture the variety of features across the images. We therefore used a variable scaling based on the flux of each target source. For each image the R, G, and B channels have a fixed ratio of $4:3:4$, and the multiplier floor was set at 2.2. 

Each colour image is 424 pixels square. The angular size of the image varies based on the size of the galaxies, according to Equations 2 and 3 of \citet[][and also see \citeauthor{kartaltepe15} \citeyear{kartaltepe15} for further details]{haeussler07}, with a minimum of $30 \times 30$ native WFC3 pixels zoomed to $424 \times 424$ display pixels. The Galaxy Zoo interface loads the normal colour images by default, and the classifier may choose to display an inverted colour image, but may not otherwise change the image scaling or size within the software while performing the classification. Classifiers are also not able to discuss galaxies before providing their classification, or pick specific galaxies to classify. This design ensures a consistent set of independent classifications which can be combined as described below.

\subsection{Photometry}

The selection of galaxies to include in Galaxy Zoo CANDELS was based on preliminary photometry of the ACS and WFC3 images, computed using Source Extractor \citep{bertin96}. As described in Section \ref{sec:images}, the sample was selected using $H \leq 25.5$~mag. 

Subsequent analysis has produced more refined photometry in each field (GOODS-S, \citeauthor{guo13} \citeyear{guo13}; UDS, \citeauthor{galametz13} \citeyear{galametz13}; COSMOS, H. Nayyeri et al., in preparation). In particular, an adapted form of Source Extractor has been used to more cleanly determine backgrounds and provide improved flux measurements. As a result, many source magnitudes have been revised to fainter values: the average source magnitude in the sample is fainter by $0.35$ mag. The faintest detected source in the revised catalog has a magnitude of $H = 28.3$. 

In general, the morphological quantities presented here do not rely on photometric information beyond initial identification of the sample. For example, we do not use colour, size, or redshift information to inform the raw or weighted morphologies. There is one exception: in Section \ref{sec:rawclass} we describe how an analysis of ongoing classifications led to a modification to the retirement limit of some subjects based on their classifiability as a function of surface brightness and magnitude. Thus for fainter, lower-surface brightness images the number of classifications may be lower than the average of $\sim40$ per subject. 

Otherwise, we only incorporated photometry into our analysis after the collection of classifications was complete. In particular, we use rest-frame $V$-band luminosities and $UVJ$ colours to show examples of how to use classifications in Section \ref{sec:release}, $H$-band {\sc auto} magnitude and 80-per-cent flux radius in Section \ref{sec:depth} when discussing depth corrections and classification biases, and rest-frame luminosities and photometric redshifts in the analysis of smooth disks in Section \ref{sec:result}. Rest-frame colours have been computed from the full multi-wavelength CANDELS photometry in each field using a template-based interpolation implemented in {\tt EAZY} \citep{brammer08} and the template set of \citet{muzzin13a}.

\subsection{Redshifts}\label{sec:z}

The choice to cover areas which had been investigated by previous surveys, and the high-profile nature of the CANDELS survey itself has ensured that each of the fields has considerable follow-up, providing a wealth of ancillary data. Of particular importance for our work is the availability of reliable estimates of redshift. Our approach has been, therefore, to gather spectroscopic and photometric data where possible. 

For COSMOS galaxies, we use spectroscopic redshifts from zCOSMOS \citep{lilly07} or, where this is not possible, photometric redshifts derived from the COSMOS survey itself \citep{ilbert09} and the NEWFIRM medium-band survey \citep{whitaker11}. For GOODS-South, \citet{cardamone10b} assembled photometric redshifts from deep imaging carried out by MUSYC \citep{gawiser06} and spectroscopic redshifts from a variety of sources  \citep{balestra10,vanzella08,lefevre04,cimatti02}. For UDS, we use available spectroscopic \citep{simpson12} and photometric redshifts \citep{hartley13}. The latter makes use of deep multi-wavelength coverage from UKIDSS as well as $J$ and $H$-band magnitudes from CANDELS itself.

Of the 49,555 galaxies originally included in Galaxy Zoo CANDELS, 46,234 currently have measured spectroscopic (2,886) or photometric (43,348) redshifts. Where available, agreement between spectroscopic and photometric redshift is generally very good, with $\Delta z \equiv \sigma_z/(1+z_{spec}) = 0.02$ and $\sim 8\%$ of sources having $\Delta z > 0.2$. The use of photometric redshifts introduces an uncertainty of less than $1\%$ into the analysis described here \citep{simmons14}. For the remaining $\sim 3000$ galaxies, we rely on photometric redshifts derived by \citet{dahlen13} who use a Bayesian approach which combines results from several different and independent approaches.

%%%%%%%%%%%%%%%%%%%%%%%%%%%%%%%%%%%%%%%%%%%%%%
%
%
\section{Classification Data}\label{sec:classifications}
%
%
%%%%%%%%%%%%%%%%%%%%%%%%%%%%%%%%%%%%%%%%%%%%%%

\subsection{Definition of Terms}\label{sec:terms}

Throughout this paper we adopt the following terms to describe different parts of the Galaxy Zoo software and data (similar to \citeauthor{willett13} \citeyear{willett13} and \citeauthor{rsimpson14} \citeyear{rsimpson14}, and used generally throughout the Zooniverse citizen science platform software\footnote{\tt www.zooniverse.org}):

\begin{itemize}
\item \textbf{Classifier}. A classifier is a volunteer participating in the project.

\item \textbf{Subject}. Within the Zooniverse software, a subject is a unit of data to be classified. 
In Galaxy Zoo CANDELS, each subject consists of a colour image and an inverted copy of the colour image, with the goal of classifying 1 galaxy per subject. (For other projects this may include light curves, groups of images, video or audio files.)

\item \textbf{Classification}. Galaxy Zoo CANDELS asks the classifier to complete several tasks to fully classify each subject. A classification is a unit of data that consists of 1 complete flow through the decision tree described in Section \ref{sec:tree}.

\item \textbf{Task} and \textbf{Question}; \textbf{Response} and \textbf{Answer}. The decision tree described below is comprised of multiple tasks the classifier is asked to complete. Each task in Galaxy Zoo CANDELS consists of a single question, with 2 or more possible responses, 1 of which the classifier selects as their answer in order to move on to the next task.

\end{itemize}

\subsection{Decision Tree}\label{sec:tree}

%%%%% [FIGURE: Decision tree] %%%%%
\begin{figure*}
\includegraphics[scale=0.85]{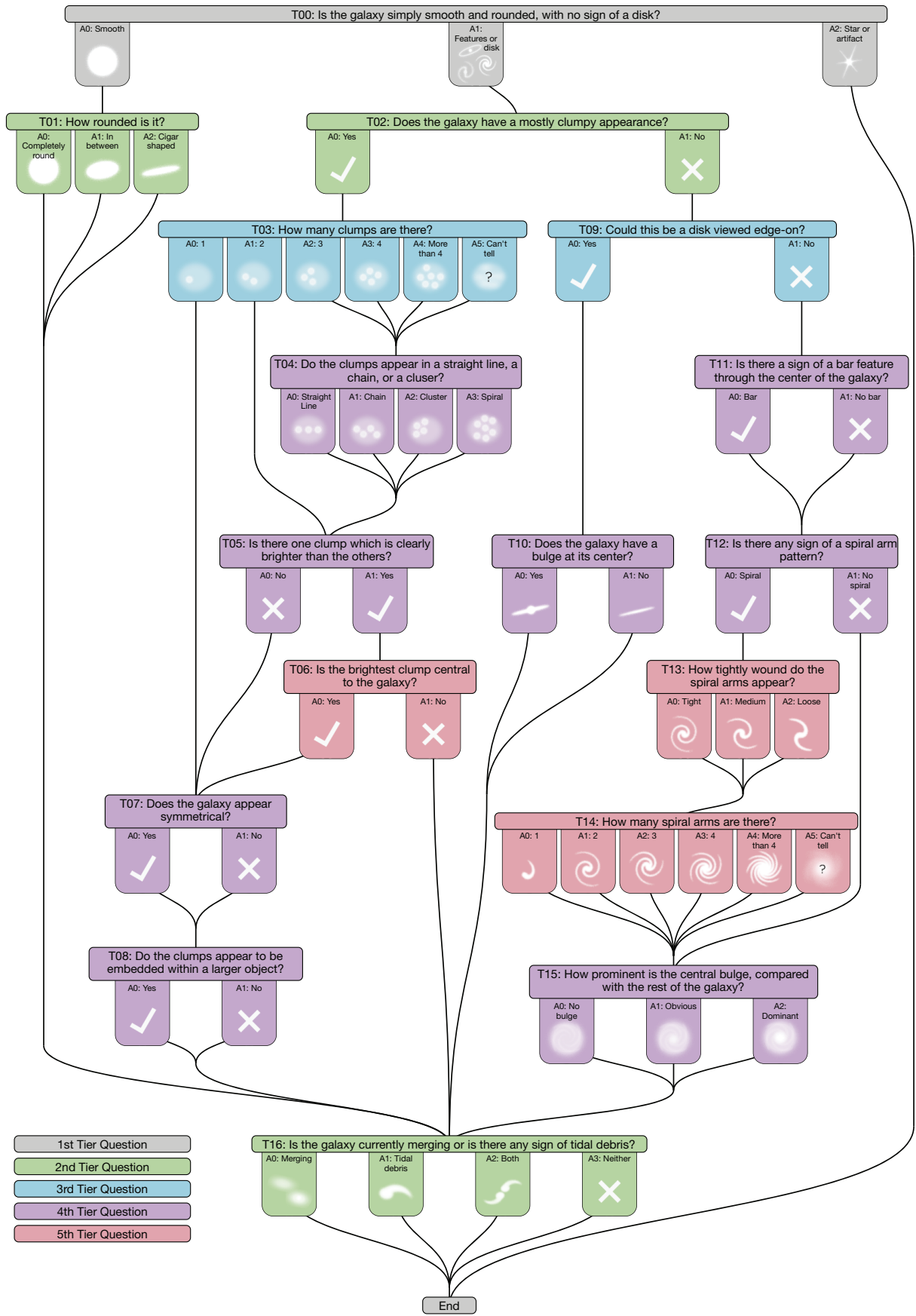}
\caption{
The decision tree for Galaxy Zoo CANDELS in visual format, including graphical icons associated with each response in the classification interface. There are 16 tasks, with one question per task and up to 6 possible responses per question. Questions are coloured according to the minimum number of branches prior to that question. All classifiers are asked the first question (task T00), and there are 4 subsequent levels of branching. The tree is also shown in text in Table \ref{table:tree}.
}
\label{fig:tree}
\end{figure*}
%%%%% END FIGURE %%%%%

The goal of Galaxy Zoo CANDELS is to provide detailed quantitative visual morphologies of galaxies observed by the deepest, most complete \emph{HST} multi-wavelength legacy survey to date. 
There are many morphological features of interest, including both broad questions about a galaxy's overall appearance and more detailed questions about specific features. 

We employ a tree-based structure for collecting information on these morphological features, a strategy that has been used successfully since Galaxy Zoo 2 \citep[GZ2;][]{willett13}. The Galaxy Zoo CANDELS decision tree is shown in visual form in Figure \ref{fig:tree} and in text form in Table \ref{table:tree}. We note that this tree is most similar to the tree used in the Galaxy Zoo: Hubble project \citep[described in][Willett et al., in preparation]{melvin14}, which also has a branch identifying clumpy galaxies and focusing on the detailed structure of galaxy clumps not present in the GZ2 tree. There are small differences between the CANDELS and GZH tree, however: for example, Task 10, which asks about a bulge in an edge-on disk, is a Yes/No question here, whereas in previous iterations of the decision tree this question also asked whether the bulge shape was rounded or boxy. Additionally, the final task in the tree (Task 16) is substantially different from previous versions and is here only concerned with galaxy mergers and tidal features. 

The CANDELS decision tree first asks the classifier to choose between the broad categories of ``smooth and rounded'', ``features or disk'', and ``star or artifact''. The next step either exits the classification (if the classifier has indicated the subject is of a star or artifact) or moves on to a task which asks for further details about the galaxy. This broad classification follows the practice of previous \emph{Galaxy Zoo} projects in making a division based solely on visual appearance, rather than attempting to infer underlying dynamical conditions. In particular, we should expect classical S0 galaxies where the disk is completely smooth and hence not easily visible in these images to be included in the ``smooth'' sample. Where such disks are visible, for example when edge-on, they will naturally appear in the ``featured'' sample.

If the classifier has indicated in the first task that the galaxy has features or a disk, the subsequent tasks ask a series of follow-up questions about features such as clumps, spiral patterns, bulge strength, and the presence of a bar. If the classifier has instead indicated the galaxy is mostly smooth and rounded, the next task asks them to rate the overall roundedness, a question roughly corresponding to an axis-ratio measurement. Finally, when the classifier has finished answering all follow-up questions about either the ``smooth'' or ``featured'' galaxy, the final task asks whether the galaxy is undergoing a merger, has tidal tails, or has both, or neither.

The tree-based structure has a number of advantages. First, it collects substantially more information on each galaxy than a single question would, and captures a more detailed classification of higher-order structures while minimising the effort required on the part of the classifier by only asking for relevant inputs based on the answers provided to previous questions. 

Second, it focuses the classifier on a single feature at a time, highlighting each feature. This resets the attention of the classifier with each new question and avoids the problems that may result when a person is presented with a large number of decision tasks at once, including a decrease in optimal decision-making \citep{iyengar00,crescenzi13, besedes15} and a reduced ability to recognise the unexpected \citep{simons99,todd05}. 

Third, the tree-based structure is especially optimal for an interface which may collect classifications from classifiers who have never before seen an image of a galaxy and may seek additional training. Within the interface, the classifier may optionally display training images in a ``help'' section that shows different examples of the feature relevant to the current question. Asking single-topic questions in turn permits a full set of training images to be available throughout the classification without placing an unnecessary cognitive load on the classifier.

The disadvantage of a tree-based classification structure concerns the dependencies introduced into the vote fractions by such a structure. A classifier cannot, for example, answer that the same galaxy has both a mostly smooth appearance and also has a bar feature. This is in some ways an advantage, as it prevents contradictory and unphysical classifications, but it also means that an analysis of morphological vote fractions with the goal of examining spiral galaxies (for example) must account for the fact that whether a given classifier reached the spiral branch of the decision tree depends on their answer to the questions preceding it. 

Accounting for dependencies of questions in deeper branches of the decision tree on higher-level questions is, however, a manageable task which has been undertaken successfully in many previous studies of specific galaxy structural features \citep[for specific examples, see e.g.][]{masters11a,melvin14,galloway15}. We provide guidelines for optimal morphological selection of samples using Galaxy Zoo consensus classifications in Section \ref{sec:usage}.

After the classification of each subject is finished, the classifier is asked ``Would you like to discuss this object?'' If the classifier selects ``no'', a new subject is shown for classification. If the classifier selects ``yes'', a new window opens with a discussion page focused on the subject they have just classified. Within this part of the Galaxy Zoo software, called Talk, people may ask questions and make comments on specific subjects, or engage in more general discussions. People may also ``tag'' subjects and discussions using a format similar to Twitter's \#hashtag system. Some of these tags were used in the pre-analysis of Galaxy Zoo CANDELS data, on which more details are given in Section \ref{sec:rawclass} below.

\begin{table*}
 \begin{tabular}{@{}cllr|cllr}
 \hline
\multicolumn{1}{l}{Task} &
\multicolumn{1}{c}{Question} &
\multicolumn{1}{c}{Responses} &
\multicolumn{1}{c}{Next} &
\multicolumn{1}{l|}{Task} &
\multicolumn{1}{c}{Question} &
\multicolumn{1}{c}{Responses} &
\multicolumn{1}{c}{Next} 
\\ 
\hline
\hline						
T00  & {\it Is the galaxy simply smooth       } & smooth            & 01         &  T09  & {\it Could this be a disk viewed   }  & yes              & 10 \\
     & {\it and rounded, with no sign of      } & features or disk  & 02         &       & {\it edge-on?                      }  & no               & 11 \\
     & {\it a disk?                           } & star or artifact  & {\bf end}  &       &                                       &                  &    \\
     &                                          &                   &            &  T10  & {\it Does the galaxy have a        }  & yes              & 16 \\
T01  & {\it How rounded is it?                } & completely round  & 16         &       & {\it bulge at its centre?          }  & no               & 16 \\
     & {\it                                   } & in between        & 16         &       &                                       &                  &    \\
     & {\it                                   } & cigar-shaped      & 16         &  T11  & {\it Is there a sign of a bar      }  & bar              & 12 \\
     &                                          &                   &            &       & {\it feature through the centre    }  & no bar           & 12 \\
T02  & {\it Does the galaxy have a            } & yes               & 03         &       & {\it of the galaxy?                }  &                  &    \\
     & {\it mostly clumpy appearance?         } & no                & 09         &       &                                       &                  &    \\
     &                                          &                   &            &  T12  & {\it Is there any sign of a        }  & spiral           & 13 \\
T03  & {\it How many clumps                   } & 1                 & 07         &       & {\it spiral arm pattern?           }  & no spiral        & 15 \\
     & {\it are there?                        } & 2                 & 05         &       &                                       &                  &    \\
     & {\it                                   } & 3                 & 04         &  T13  & {\it How tightly wound do the      }  & tight            & 14 \\
     & {\it                                   } & 4                 & 04         &       & {\it spiral arms appear?           }  & medium           & 14 \\
     & {\it                                   } & more than four    & 04         &       & {\it                               }  & loose            & 14 \\
     & {\it                                   } & can't tell        & 04         &       &                                       &                  &    \\
     &                                          &                   &            &  T14  & {\it How many spiral arms          }  & 1                & 15 \\
T04  & {\it Do the clumps appear in           } & straight line           & 05         &       & {\it are there?                    }  & 2                & 15 \\
     & {\it a straight line, a chain          } & chain            & 05         &       & {\it                               }  & 3                & 15 \\
     & {\it or a cluster?                     } & cluster     & 05         &       & {\it                               }  & 4                & 15 \\
     & {\it                                   } & spiral             & 05         &       & {\it                               }  & more than four   & 15 \\
     &                                          &                   &            &       & {\it                               }  & can't tell       & 15 \\
T05  & {\it Is there one clump which is       } & yes               & 06         &       &                                       &                  &    \\
     & {\it clearly brighter than the others? } & no                & 07         &  T15  & {\it How prominent is the          }  & no bulge         & 16 \\
     &                                          &                   &            &       & {\it central bulge, compared       }  & just noticeable  & 16 \\
T06  & {\it Is the brightest clump            } & yes               & 07         &       & {\it with the rest of the galaxy?  }  & obvious          & 16 \\
     & {\it central to the galaxy?            } & no                & 16         &       & {\it                               }  & dominant         & 16 \\
     &                                          &                   &            &       &                                       &                  &    \\ 
T07  & {\it Does the galaxy                   } & yes               & 08         &  T16  & {\it Is the galaxy currently       }  & merging          & {\bf end}  \\
     & {\it appear symmetrical?               } & no                & 08         &       & {\it merging or is there any       }  & tidal debris     & {\bf end}  \\
     &                                          &                   &            &       & {\it sign of tidal debris?         }  & both             & {\bf end}  \\
T08  & {\it Do the clumps appear to be        } & yes               & 16         &       & {\it                               }  & neither          & {\bf end}  \\
     & {\it embedded within a larger object?  } & no                & 16         &       &                                       &                  &    \\
 \end{tabular}
 \caption{The Galaxy Zoo CANDELS decision tree, comprising 16 tasks and 51 responses. Each task is comprised of a single question and up to 6 possible responses. The first question is Task 00, and a classification is completed by responding to all subsequent questions until the end of the tree is reached. The `Next' column indicates the subsequent task the classifier is directed to upon choosing a specific response. Although a classifier will flow through the tree from top to bottom, there is no path through the tree that includes all tasks. 
\label{table:tree}}
\end{table*}

\subsection{Raw classifications}\label{sec:rawclass}

% First classification 2012-09-10 18:41:25 UTC, last classification 2013-11-30 12:58:51 UTC

%%%%% [FIGURE: Classification counts per subject and user] %%%%%
\begin{figure*}
\includegraphics[scale=0.87]{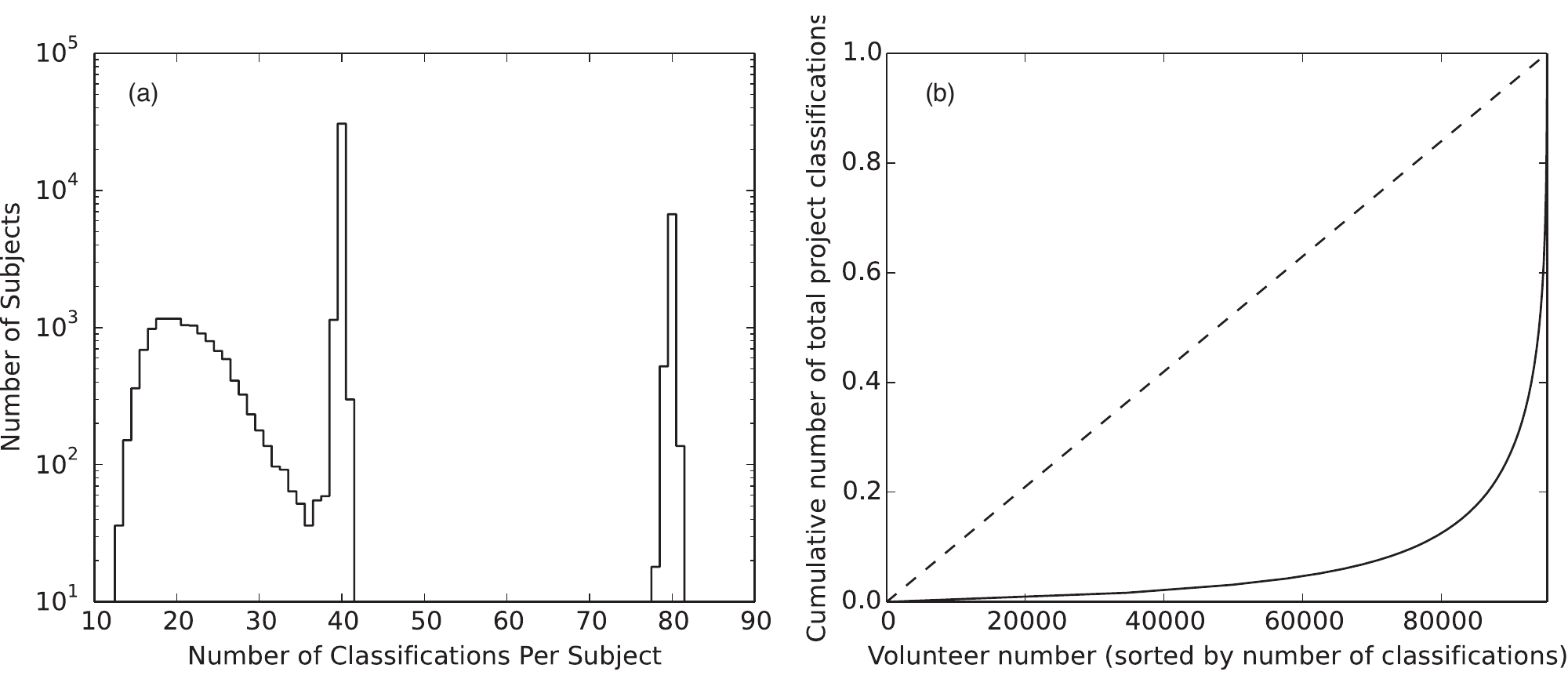}
\caption{
Basic information on classifications. \emph{Left}: Distribution of (unweighted) number of classifications per subject in Galaxy Zoo CANDELS. The majority of subjects have 40 independent classifications each; a subset of 13,392 were retired early after being identified as too faint and low-surface brightness for additional classifications to be useful (11,837) or as stars or artifacts (1,555). Subsequently, 7,402 subjects where at least 20\% of classifiers registered a vote for ``features or disk'' in the first task were re-activated with a retirement limit of 80 classifications, in order to ensure a complete sampling of the deepest branches of the decision tree. \emph{Right}: Cumulative distribution of classifications per classifier, where the classifiers are sorted in order of least to most classifications contributed (Lorenz curve for classifiers). If every classifier had contributed the same number of classifications, the Lorenz curve would be equal to the dashed curve. The top 9\% of classifiers contributed 80\% of the classifications (Gini coefficient = 0.86). 
}
\label{fig:classification_basic_info}
\end{figure*}
%%%%% END FIGURE %%%%%

The first classification of a subject from CANDELS was registered on the Galaxy Zoo interface\footnote{zoo4.galaxyzoo.org} on the 10th of September 2012. The final classification considered here, in the first phase of Galaxy Zoo CANDELS, was registered on the 30th of November 2013. Between these times, the site collected 2,149,206 classifications of 52,073 CANDELS subjects (of which 2,518 were intentional duplicates of the same galaxy; see Section \ref{sec:depth}) from 41,552 registered classifiers and 53,714 web browser sessions where the classifier did not log in. For all analysis presented here we have assumed that each unregistered browser session contains classifications from a single, unique classifier. 

Subjects within a given Galaxy Zoo sample are chosen randomly for classification, so that the number of independent classifications per galaxy builds up uniformly through the full sample. Once a pre-set classification minimum count has been reached, the subject is retired from the active classification pool. While more sophisticated task assignment might have led to efficiency savings \citep[e.g.][]{Waterhouse13}, it would have changed the experience of what was a popular project \citep{Bowyer15} in unpredictable ways, and the infrastructure supporting the project did not at the time make such complexities possible. The initial goal for Galaxy Zoo CANDELS was thus simply to obtain at least 40 independent classifications for each galaxy. 

This uniform retirement limit was modified twice during the project. In the first instance, a pre-analysis of the dataset performed when the average number of classifications per galaxy had reached approximately 20 revealed 11,837 subjects where further classification was unlikely to provide significant additional information. These subjects were identified with the help of a set of subjects tagged in the Galaxy Zoo Talk software as ``\#toofainttoclassify'' and ``\#FHB'' (which stands for ``Faint Hubble Blob''). Tags in Galaxy Zoo Talk are generally highly incomplete; thus the 204 tagged subjects were used as tracers during a further examination of all subjects in magnitude-surface brightness parameter space. The selection, made from initial photometry, was deliberately conservative, retiring only those subjects where it was clear that the classification vote fractions had converged at all tiers of the classification tree. During this analysis, an additional 1,555 subjects were identified as highly likely to be stars or artifacts and were also retired.

The second modification of the retirement limit was implemented 1 year after the project start. At this time, the retirement limit was raised to 80 classifications for all galaxies where at least 20\% of classifiers had answered ``features or disk'' to the first question (task T00 in Figure \ref{fig:tree} and Table \ref{table:tree}). This is a higher retirement limit than in previous Galaxy Zoo projects, and it is justified by the increased complexity of the decision tree compared to, e.g., Galaxy Zoo 2 \citep{willett13}. The Galaxy Zoo CANDELS decision tree has an additional branch level, and the number of classifiers answering a question is typically reduced at each branch point. Thus, 40 classifications at the first question may not be enough to ensure convergence in, for example, task 14, ``How many spiral arms are there?'', a 5th-tier task with 6 possible responses. The increased retirement limit affected 7,402 subjects.

Figure \ref{fig:classification_basic_info}a shows the distribution of total classification counts within the sample. The majority of subjects received 40 classifications, but the distribution is asymmetric: there are peaks at $\sim 20$, $40$, and $80$ classifications, consistent with the description above. The Lorenz curve of classifications (ie. the cumulative number of classifications in order of classifier contribution) is shown in Figure \ref{fig:classification_basic_info}b. The curve is highly skewed from the $1:1$ line that would be seen if all classifiers contributed the same number of classifications; the top 9\% of classifiers contributed 80\% of total classifications. The Gini coefficient for classifications, i.e., the fractional difference in area under the Lorenz curve versus the dashed line, is 0.86. This is typical of past Galaxy Zoo projects and Zooniverse citizen research projects in general \citep{cox15}.

The values in Figure \ref{fig:classification_basic_info} are raw classification counts; while raw classification counts and vote fractions are certainly useful, and included in the data release described in Section \ref{sec:release}, we additionally ``clean'' the data with a simple method to identify seriously errant classifiers (most likely from bots), and then apply a classifier weighting scheme to classifications to produce a cleaner set of vote fractions for each subject. Both steps are described in further detail below.

\subsection{Identification and removal of single-answer prolific classifiers}\label{sec:bots}

Within the raw classifications, a small group of classifiers (86, or less than 0.1 per cent) classified at least 200 subjects \emph{and} gave the same answer to the question in the first task at least 98 per cent of the time. Within this group, 99.6 per cent of classifications were for ``star or artifact'' (from 84 classifiers) and 0.4 per cent were for ``smooth'' (from 2 classifiers). 

Only a small number of unresolved sources or sources dominated by an unresolved element (i.e., stars and quasars) are included in the full Galaxy Zoo CANDELS subject sample. Examination of the CANDELS photometric catalogs \citep[][H. Nayyeri et al., in preparation]{galametz13,guo13} shows that less than 12 per cent of subjects have $\stellarity > 0.25$ (a very inclusive cut; a more typical cut on stellarity estimates the number of unresolved sources at less than 3 per cent). If the subjects assigned to a classifier are drawn at random from the subject set, then for any classifier who submits a substantial number of classifications, the chances they will be shown a large fraction of stars is very small. 

Even for more common answers to the first task, the chances of a classifier being randomly assigned a highly uniform set of $N$ subjects becomes very small as $N$ becomes large. For example, if the probability of being assigned a ``smooth'' galaxy is $p = 0.9$, the chance of being assigned a subject set of 98 per cent smooth galaxies out of $N > 200$ total is so small that it would likely happen approximately once per billion classifiers, \emph{i.e.}, it is highly unlikely in a project with $\sim 100,000$ classifiers.

As the chances of any classifier being actually served $> 98$ per cent of subjects with the same intrinsic classification in more than 200 classifications is vanishingly small, these classifiers are most likely bots or are otherwise not actually engaging in the classification task. While these classifications ($6.8$ per cent of the total classifications) would be substantially down-weighted during the classifier weighting process described below, we formally omit them from further analysis and do not include them in the weighting and consensus calculations. The fraction of bot-like classifications in this project is consistent with that found in the first Galaxy Zoo project, where approximately $4$ per cent of classifications were removed for the same reasons \citep{lintott08}. The average number of classifications per unique subject after excluding the omitted classifications is 40.4. 

We did not manually search for classifiers whose inputs are consistent with random or otherwise suspect; these inputs, if they exist, are effectively down-weighed via the consensus-based classifier weighting described below, within which prolific classifiers tend to have very high consistency values.

\subsection{Classifier Weighting} \label{sec:weighting}

Multiple methods of classifier weighting have been successfully employed by different Zooniverse projects \citep{lintott08, bamford09, lintott11, rsimpson12, schwamb12,esimpson13, johnson15, marshall16}. In general, the optimal choice of classifier weighting depends on the amount of information available per subject and the goal of the project. In Galaxy Zoo CANDELS the goal is to converge to a classification for each galaxy whilst still allowing for unexpected discoveries. There is ample information from classifiers but little information on the ``ground truth'', i.e., we do not know what the true intrinsic classification is for even a modest fraction of the sample.

For these reasons, we apply a consensus-based weighting method for the majority of the tasks in the decision tree, informed first by the application of initial weights based on comparison of the classifications in the first task to the stellarity (\stellarity ) parameter from the CANDELS photometric catalogs. Both are described below, in the order in which they are applied.

\subsubsection{Initial weighting based on ``star'' versus ``galaxy'' classifications}

In the initial classification task (T00), we ask classifiers to separate stars from galaxies and identify a galaxy as ``smooth'' or as having ``features or disk''. Although we have no ``ground truth'' information on the overall morphology of a galaxy, we do have very reliable information on whether the source detected in each image is extended, from the \stellarity\ parameter. We can therefore apply classifier weightings to this task based on whether classifiers typically classify bright stars as ``star or artifact'', and whether they classify extended objects as galaxies (i.e., whether they answer \emph{either} ``smooth'' or ``features or disk'').

We select a sample of bright stars having $F160W < 18.5$ and $\stellarity  > 0.8$ from within the Galaxy Zoo-CANDELS subject set. After manually rejecting 2 subjects which contain a galaxy in the central image position with a bright star nearby or overlapping, the bright-star gold-standard sample contains 263 subjects.

We select a sample of extended sources having $F160W < 25$ and $\stellarity < 0.03$, with further manual removal of images with artifacts and other ``unclassifiable'' sources. We first cleaned this sample by rejecting remaining sources where more than 65 per cent of classifiers had selected the ``star or artifact'' response to task T00, a choice made to favour purity of the extended-source sample over completeness. We additionally rejected 398 artifacts falling below this threshold, leaving a total of 29,996 subjects in the extended-source gold-standard sample. 

Having selected these samples, we then assigned an index $n_s$ to each subject classification from within either gold-standard subject set. For subjects within the bright-star gold-standard set, the classification index was set to $n_s = -1$ if the classifier had \emph{not} marked the subject as ``star or artifact'', and was $n_s = 0$ otherwise. For subjects within the extended-source gold-standard set, the classification index was set to $n_s = -1$ if the classifier had marked the subject as ``star or artifact'', and was set to $n_s = +1$ otherwise.

We then define the index $n_c$ for each classifier as the sum of all their classification indices $n_s$, and the weight for task T00 is assigned based on the classifier index as
\begin{equation}
   w_{00}= \left\{
    \begin{array}{l l}
      \max \left(1.1^{n_c}, 0.01\right)       & \text{ if } n_c < 0 \\
      \min \left(1.05^{n_c}, 3 \right) & \text{ if } n_c \geq 0.\\
    \end{array} \right.
    \label{eqn-seedweight}
 \end{equation}

This weighting results in a set of classifier weights between $0.01 < w_{00} < 3$, with classifiers whose classifications are generally ``correct'' being up-weighted and classifiers who are more often ``incorrect'' being down-weighted.  79 per cent of classifiers classified at least 1 subject within either gold-standard subject set; classifiers who did not classify any subjects in the gold-standard subject set have $w_{00} = 1$. Of the classifiers who were included in the weighting, 56 per cent have $w_{00} > 1$, with a mean of $\left< w_{00} \right> = 1.11$. As a last step, the weights are re-normalised so that the sum of weights is equal to the total number of classifications.

Following this initial weighting, we create an initial set of vote fractions for each subject by summing the weighted votes for each task and response, and reporting the vote fractions $f$ for each. We use this as an initial consensus classification catalog in the consensus-based weighting applied to the remaining tasks, described in further detail below.

\subsubsection{Consensus-based classifier weighting}

Following the weighting of task T00 described above, we adopt an iterative consensus-based weighting method for classification tasks T01 through T16. This weighting scheme follows previous Galaxy Zoo projects and effectively identifies the small proportion of classifiers whose contributions are routinely errant compared to other classifiers (or consistent with random inputs) and downweights their contributions, while preserving the inputs from the vast majority of classifiers.

Weights for each classifier are computed based on a mean consistency factor, \kappamean , which is the average of consistencies for each of that classifier's classifications. For a given classification $i$ composed of a series of completed tasks $t$ answered about a specific subject, we compare the classifier's answer to each task with the aggregated classifications of all classifiers of the same subject. Each task has $a_t$ answers from all classifiers, each of which is assigned to one of $N_{r,t}$ possible responses to the task. We define the vote fraction for a particular response $r$ as $f_r \equiv a_r/a_t$, where $a_r$ is the number of positive answers for that response (i.e., the number of classifiers who selected that response out of all possible responses to the task).

For each task that was completed by the classifier in classification $i$, the consistency index $\kappa_r$ for each response $r$ to that task $t$ is 
\begin{equation}
    \kappa_r = \left\{
    \begin{array}{l l}
      f_r       & \text{ if the classifier's answer corresponds} \\
                  & \text{ to this response,}\\
      (1 - f_r) & \text{ if the answer does not correspond.}\\
    \end{array} \right.
    \label{eqn-consistency-r}
 \end{equation}
The consistency for that task, $\kappa_t$, is the average of these indices over all possible responses. For example, if a classifier answered ``star or artifact'' to Task T00 for a particular subject, and the overall vote fractions on that task for that subject are $($``smooth'', ``features or disk'', ``star or artifact''$) = (0.1, 0.6, 0.3)$, then the classifier's consistency for Task T00 for this classification is
$$
\kappa_t = \left[\left(1 - 0.1\right) + \left(1 - 0.6\right) + 0.3\right]/3 = 0.5\overline{3}.
$$
In the above example, the classifier's answer to Task T00 leads to the end of the workflow (Table \ref{table:tree}), so this $\kappa_t$ is also equal to the classifier's consistency for the overall classification, $\kappa_i$. More generally, the classification consistency is the answer-weighted average of the task consistencies:
\begin{equation}
\kappa_i = \frac{\sum\limits_t \kappa_t a_t}{\sum\limits_t a_t} ,
\label{eqn-consistency-i}
\end{equation}
where each sum is over the number of tasks the classifier completed during the classification.

Following this calculation for the entire classification database, each classifier's average consistency is calculated as
\begin{equation}
\mkappamean = \frac{1}{N_i} \sum\limits_i \kappa_i .
\label{eqn-consistency-avg}
\end{equation}

Averaging over a classifier's individual consistency values for all classifications effectively downweights those contributions from classifiers whose classifications regularly diverge from the consensus whilst preserving the diversity of classifications from classifiers who are \emph{on average} consistent with each other. It also allows for the classifications of skilled classifiers to remain highly weighted even on difficult subjects where the individual consensus is skewed (e.g., if an image is very noisy or if a nearby artifact is distracting to less experienced classifiers). 

The classifier weight is then calculated as 
\begin{equation}
w = \min \left(1.0,(\mkappamean / 0.6)^{8.5} \right) ,
\label{eqn-weight}
\end{equation}
a formulation that preserves a uniform weighting for any classifier with $\mkappamean \geq 0.6$ and downweights those with a lower consistency rating.

The weighted consensus classifications are then calculated for each subject by summing the weighted votes for each task and response between task T01 and T16, and reporting the vote fractions $f$ for each. (Although the classifications for task T00 are included in the computation of the consensus-based weights, the vote fractions for task T00 are not re-computed using the consensus-based weights.) As the classifier weights are calculated via comparison with the consensus, which leads to a new consensus, this method can be iterated until the classifier weights converge to a stable value. 

In practice, the number of iterations required to reach this goal is low \citep[e.g., 3 or less in previous projects;][]{bamford09,willett13}. In Figure \ref{fig:consistencies} we show the distribution of classifier consistencies after 1-5 iterations of the above method, although the difference between iterations 3 through 5 cannot be distinguished within the line weight even in the inset (zoomed) subsection of the figure. Between the 4th and 5th iterations, more than 99 per cent of consistency values varied by less than 0.1 per cent. After 5 iterations, approximately 4 per cent of classifiers have consistency $\mkappamean < 0.5$ (corresponding to a weight $w \lesssim 0.2$), whereas 83 per cent of classifiers have an end weight of $w = 1$. The vast majority of Galaxy Zoo classifiers thus contribute highly valuable information to the project.

Figure \ref{fig:exampleclass} shows examples of galaxies with different weighted consensus classifications for several of the tasks described in Table \ref{table:tree} and Figure \ref{fig:tree}. Figure \ref{fig:sankey} shows the demographics of the full sample, using the weighted vote fractions to assign a single label to each galaxy at each task in the classification tree. We only consider galaxies for a given task if they were assigned an appropriate label that flows into that task (Figure \ref{fig:tree}). Galaxies are considered ``featured'' if at least 30\% of (weighted) classifiers answered``features or disk'' in the initial task; for all other tasks and responses we assign labels based on a plurality of weighted responses. 

This very simple method provides an overview of the sample demographics. The majority of galaxies in the sample would be considered ``smooth'', with only 12\% labelled as ``featured'' even given the relatively generous selection of featured galaxies. Were we to select subsets of galaxies based on other criteria, such as stellar mass, the demographics would likely change substantially. For example, the vast majority of the galaxies that were retired early based on their low surface brightnesses and sizes (Section \ref{sec:rawclass}) would be considered smooth or, owing to the high noise levels relative to the detected galaxy, labelled as artifacts. 

%%%%% [FIGURE: classifier Consistencies] %%%%%
\begin{figure}
\includegraphics[scale=0.85]{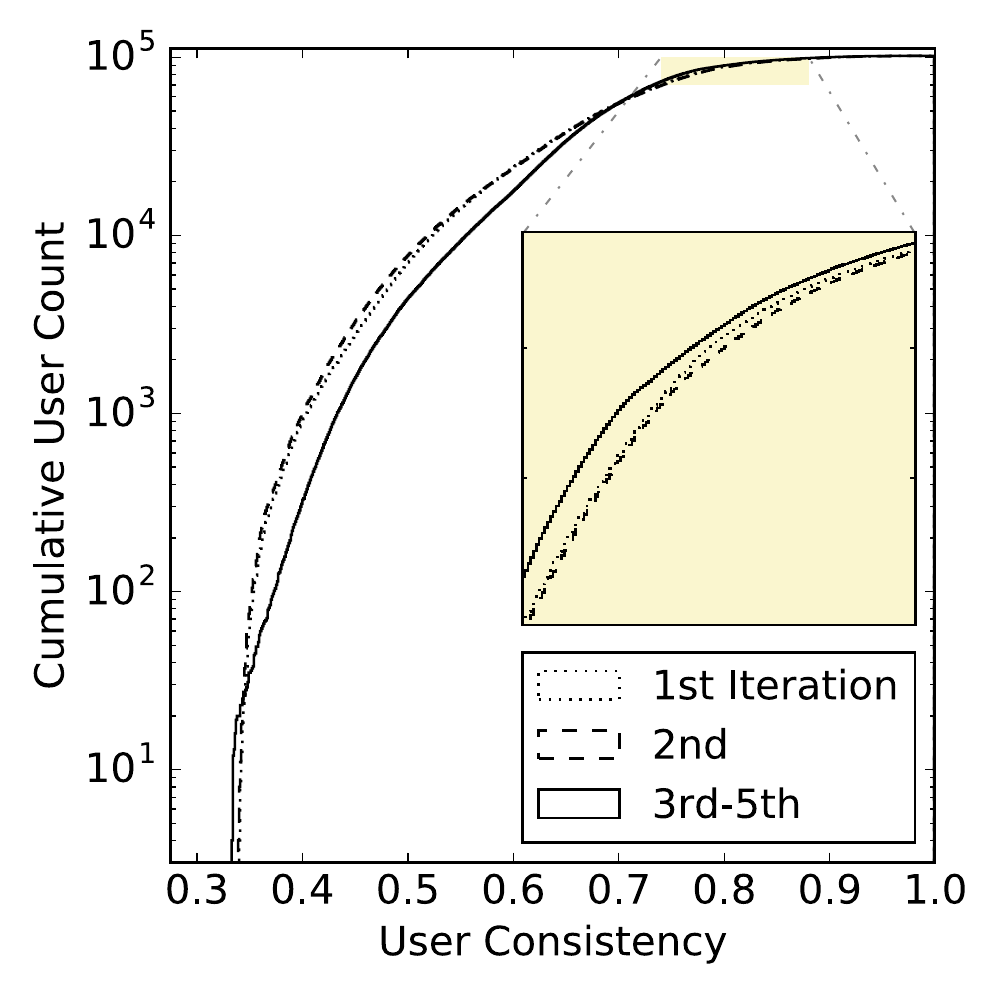}
\caption{
Distribution of classifier consistencies \kappamean\ after 1 (black dashed), 2 (black dotted), and 3 or more (black solid) iterations of the consistency-based weighting method (described in Section \ref{sec:weighting}). A portion of the plot is magnified (inset) to show further detail. Convergence of this method requires relatively few iterations, consistent with previous Galaxy Zoo projects. Approximately 83 per cent of classifiers have $\mkappamean \geq 0.6$ and weights $w = 1$. 
}
\label{fig:consistencies}
\end{figure}
%%%%% END FIGURE %%%%%

%%%%% [FIGURE: Example images] %%%%%
\begin{figure*}
\includegraphics[scale=0.875]{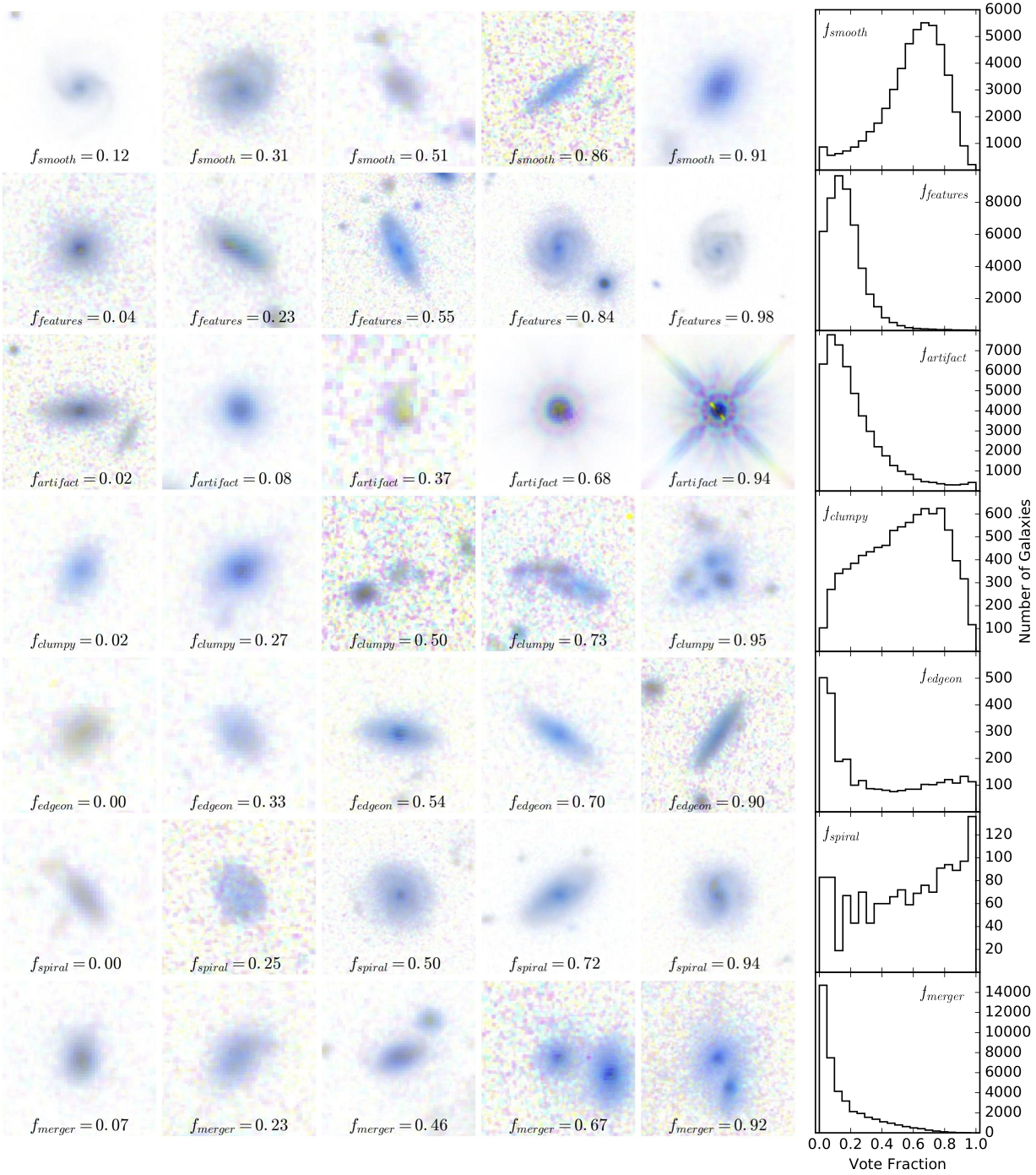}
\caption{
Example (inverted) galaxy images for different consensus classifications of different responses to tasks in the Galaxy Zoo CANDELS classification tree. From top to bottom row, the responses are: Task T00, ``smooth''; Task T00, ``features or disk''; Task T00, ``star or artifact''; Task T02, ``yes'' (Clumpy); Task T09, ``yes'' (Edge-on); Task T12, ``yes'' (Spiral); Task T16, ``merging''. Each image is labelled with the weighted percentage of total votes for that task that were registered for that response, with the weighted vote percentage increasing from left to right. The galaxies were selected after following the suggestions in Section \ref{sec:usage} regarding selection of appropriate samples, including restrictions on votes from earlier branches in the tree (see Figure \ref{fig:tree} for more information on branches). The right-most column shows vote fraction distributions for the task and response in each row, among galaxies where at least 10 answers total were received for that question.
}
\label{fig:exampleclass}
\end{figure*}
%%%%% END FIGURE %%%%%

%%%%% [FIGURE: Sankey diagrams] %%%%%
\begin{figure*}
\hfill
\includegraphics[width=8.5cm]{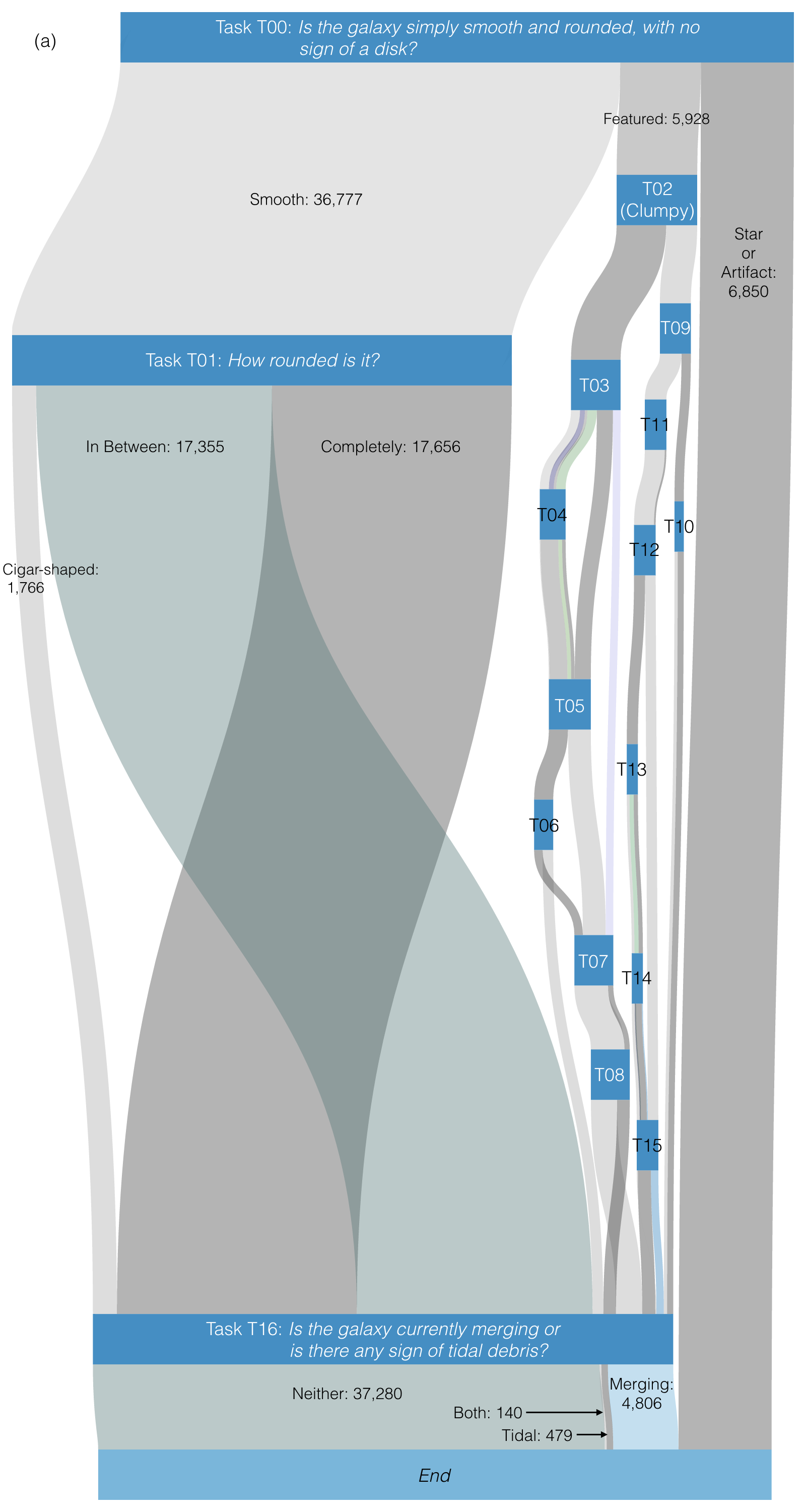}
\hfill
\includegraphics[width=8.5cm]{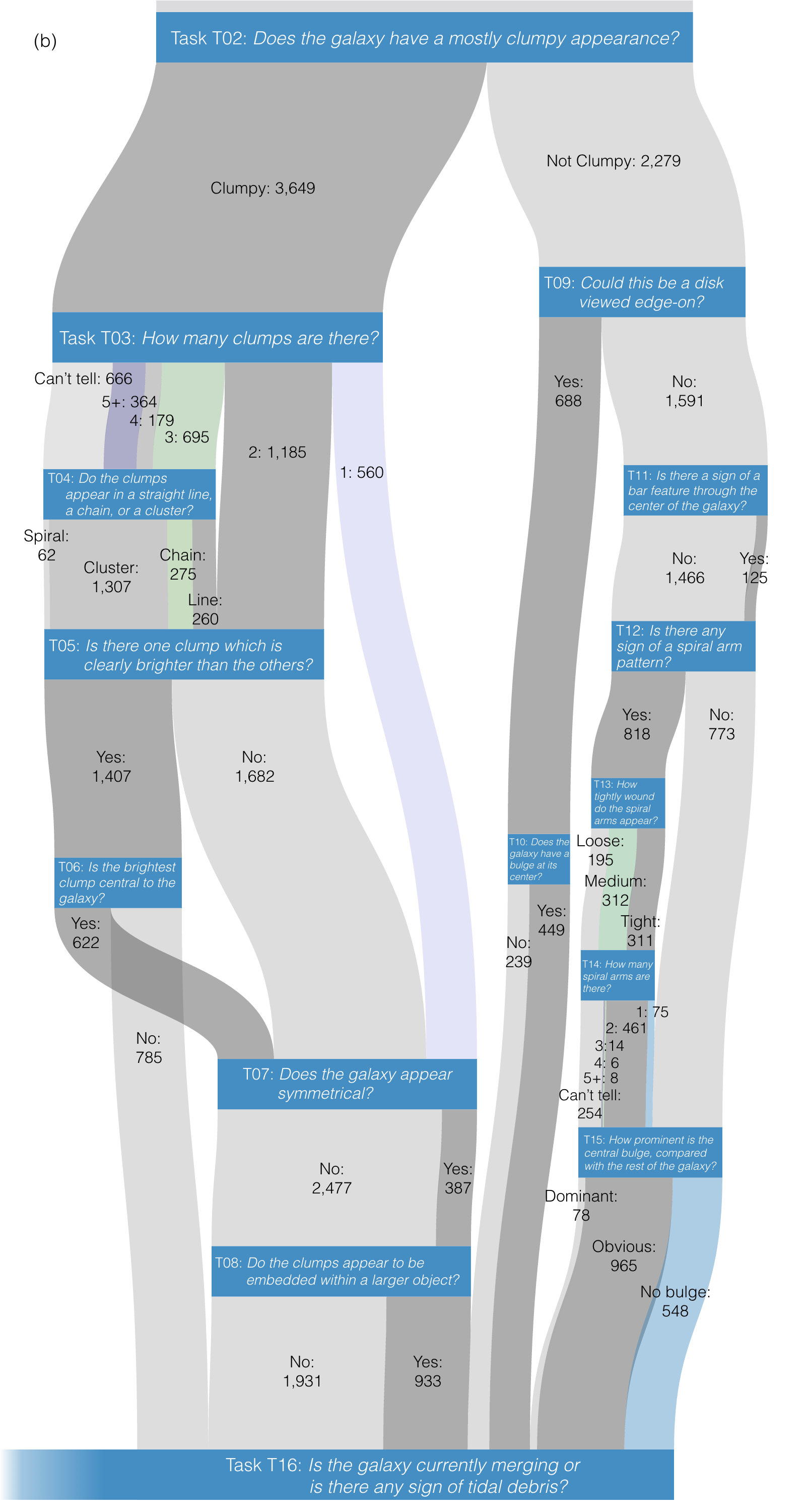}
\hfill
\caption{
Demographics of the Galaxy Zoo CANDELS sample, using the aggregate distribution of weighted morphologies. In the left panel, the full classification tree (Figure \ref{fig:tree} and Table \ref{table:tree}) is shown. The right panel shows only tasks T02-T16, the ``featured'' branches of the tree, for which the full detail is difficult to see in the left panel. Each node in each diagram (dark blue horizontal bars of uniform height) represents a task in the tree. The paths between tasks represent each possible answer to the task; these flow from top to bottom between their origin question and the subsequent task in the tree. For the first task, T00, a galaxy is considered ``featured'' if the weighted vote fractions meet the criteria $\left(\mffeat \geq 0.3\ \&\ f_{\rm star\ or\ artifact} < 0.5\right)$. From among galaxies not considered ``featured'', a galaxy is then assigned a label (i.e., to a path) of ``smooth'' or ``artifact'' based on the plurality classification between those 2 responses. For the remaining tasks, the label for an eligible galaxy is assigned according to the plurality answer for that task. At each node, a galaxy is assigned only one label. The widths of the paths are proportional to the number of galaxies assigned to that path; the widths of the nodes are proportional to the number of galaxies for which the question was reliably answered. The plurality answer represents the single most common response to a task --- this may be either above or below 50\%, depending on the number of answers and level of consensus. While this method provides a useful overview of the morphologies of the entire data set, we note that examining subsets based on various cuts (for example, stellar mass or luminosity thresholds) may reveal very different demographics for those sub-populations.
}
\label{fig:sankey}
\end{figure*}
%%%%% END FIGURE %%%%%

\subsection{Use of Classifications in Practice}\label{sec:usage}

The branched nature of the decision tree (Figure \ref{fig:tree}) means that selection of a sample of galaxies for a given morphological investigation may depend on a number of factors. For example, it is possible to choose a quantitative threshold for selection of a sample of galaxies with a given feature or combination of features corresponding to one's optimal trade-off between sample completeness and purity. One may also weight a population analysis by the vote fraction for a particular morphological feature \citep[making the assumption that the vote fraction is a well-behaved estimator of the true probability of a galaxy having the relevant feature, e.g.,][]{skibba09,smethurst15}. However, for all tasks below T00 in the tree, it is important to consider the responses to the tasks above it in this analysis.

For example, a study with the goal of examining spiral galaxies would ideally use a sample selected by considering the responses to task T12, ``Is there any sign of a spiral arm pattern?'' If a pure sample of galaxies with clear spiral arms is desired, a threshold may be selected at a high vote fraction for $f_{\rm spiral}$. If the threshold considers only this vote fraction, however, the final sample will likely be contaminated by galaxies where the spiral vote fraction is dominated by noise because only a small number of people reached that task (e.g., a warped edge-on disk). 

In order to reach task T12, a classifier must give specific answers to the questions ``Is the galaxy simply smooth and rounded, with no sign of a disk?'' (T00), ``Does the galaxy have a mostly clumpy appearance?'' (T02), and ``Could this be a disk viewed edge-on?'' (T09). Each of these classifications should be considered in the context of this hypothetical study's goals in order to select as pure a sample as possible whilst minimising contamination and bias.

If a moderately complete sample is desired, for example, the user could select thresholds for the selection such as $\mffeat > 0.5$, $f_{\rm not\ clumpy} > 0.5$, $f_{\rm not\ edge-on} > 0.5$. Because most galaxies with these classifications will have received 80 classifications apiece (Section \ref{sec:rawclass}), these chained thresholds mean the minimum number of classifiers who will have answered the spiral question for subjects that are included in the sample is $80 \times 0.5^3 = 10$. Higher thresholds will further increase the minimum number of respondents to the deeper-branched task. If lower thresholds are desired, we recommend that the selection explicitly require a minimum number of respondents to each task. 

There is no single set of thresholds that is ideal for all situations. However, in the data release accompanying this paper, we include ``clean'' selections of galaxies with different morphological features. These are detailed further below.

\subsection{Data release and ``clean'' samples}\label{sec:release}

This paper includes the release of the raw and weighted classifications for each of the $49,555$ subjects in the Galaxy Zoo CANDELS sample. In addition to each raw and weighted vote fraction for each task, we include the raw and weighted number of answers to each task, as well as the total raw and weighted classifier count for each subject. This combines for a total of 136 quantities for each subject, not including the subject ID or any other metadata. The structure of the data for each task number $NN$ with $i = {0 {\rm\ to\ } n-1}$ responses is as follows:
%oh god it's really late and I just said "thusly"

% THUSLY REMOVED - KWW

%THUSLY CERTIFIED 'THUSLY FREE' - CJL

\begin{itemize}
\item[] \small{\tt t[$NN$]\_[quest\_abbrev]\_a[$i$]\_[resp\_abbrev]\_frac} : the raw fraction of classifiers who gave this response. {\tt quest\_abbrev} and {\tt resp\_abbrev} are abbreviated versions of the specific question and response, respectively.

\item[] \small{\tt t[$NN$]\_[quest\_abbrev]\_a[$i$]\_[resp\_abbrev]\_weighted\_frac} : the weighted fraction of classifiers who gave this response. 

\item[] \small{\tt t[$NN$]\_[quest\_abbrev]\_count} : the raw count of classifiers who responded to this task. 

\item[] \small{\tt t[$NN$]\_[quest\_abbrev]\_weight} : the weighted count of classifiers who responded to this task. 

\end{itemize}

For example, the information available for task T00 (which has 3 responses) is structured as:

\begin{itemize}
\item[] \small{\tt t00\_smooth\_or\_featured\_a0\_smooth\_frac}

\item[] \small{\tt t00\_smooth\_or\_featured\_a1\_features\_frac}

\item[] \small{\tt t00\_smooth\_or\_featured\_a2\_star\_or\_artifact\_frac}

\item[] \small{\tt t00\_smooth\_or\_featured\_a0\_smooth\_weighted\_frac}

\item[] \small{\tt t00\_smooth\_or\_featured\_a1\_features\_weighted\_frac}

\item[] \small{\tt t00\_smooth\_or\_featured\_a2\_star\_or\_artifact\_weighted\_frac} 

\item[] \small{\tt t00\_smooth\_or\_featured\_count} 

\item[] \small{\tt t00\_smooth\_or\_featured\_weight}

\end{itemize}

The sum of raw  {\tt \_frac} fractions adds to 1.0, as does the sum of {\tt \_weighted\_frac} fractions. Multiplying the {\tt \_frac} values (raw fractions) by the {\tt \_count} (raw classifier counts) will retrieve the number of people who gave a specific answer; likewise with weighted answer counts from {\tt \_weighted\_frac} and {\tt \_weight}. As the consensus-based classifier weighting described in Section \ref{sec:weighting} assigns a weight of $w \leq 1$ to each classifier, the weighted vote count for tasks T01--T16 must be less than or equal to the raw vote count for those tasks. While the raw vote counts and fractions are provided for completeness, we recommend that users of this data set use the weighted fractions and counts. The raw and weighted classifications are presented in Table \ref{table:data-main}.

In addition to the vote fractions for each subject, we provide a set of flags for each subject that indicates its member or non-member status in a ``clean'' sample of galaxies of a specific type. We select separate clean samples of smooth, featured, clumpy, edge-on, and spiral galaxies. These samples contain exemplars of each galaxy type with minimal contamination of the sample --- as a result, samples selected with the flags will be highly incomplete, but also highly pure. They are selected according to vote fraction and vote count thresholds given in Table \ref{table:clean}, which were selected following visual inspection of the resulting samples. The thresholds roughly coincide with advice given in \citet{kartaltepe15}, who consider classifications for systems with $\mathrm{H} > 24.5$ unreliable; only 8 systems in the ``clean'' spiral sample (2\% of the sample) are fainter than that limit.

We provide the ``clean'' flags for the convenience of the end user, but we additionally encourage those wishing to use Galaxy Zoo classifications to investigate whether a different set of thresholds would be optimal for their own science case. For example, it should be noted that the redshift and magnitude distributions of samples constructed using these thresholds will be different from the main sample. We show in Figure \ref{fig:Mz_thresh} the luminosity-redshift distributions of galaxies identified as smooth and featured with a range of different thresholds; brighter, nearby galaxies are clearly more likely to appear in featured samples based on more restrictive thresholds. \citet{bamford09} were able to adjust for this bias in classifications of SDSS galaxies by assuming that no evolution took place across the sample, an assumption which is clearly not valid here. The study of Galaxy Zoo classifications in other Hubble surveys by Willett et al. (2016, in prep) used artificially redshifted images to quantify this bias, but no such images are yet available for the range of redshifts probed by CANDELS. 

Figure \ref{fig:Mz_thresh} shows that the peak of the distribution of rest-frame $V$-band absolute magnitudes in a sample of featured galaxies varies from $M_{V,~\rm peak} = -21.1$ when the sample is selected to have $\mffeat \geq 0.4$, to $M_{V,~\rm peak} = -22.0$ when the sample is selected to have $\mffeat \geq 0.7$ (the ``clean'' featured sample); the full sample of all galaxies has $M_{V,~\rm peak} = -19.9$. The peak redshift of featured samples is relatively insensitive to the choice of threshold, for threshold values of at least 0.4. However, more restrictive thresholds progressively remove a tail of galaxies with higher redshifts. For example, 35 per cent of galaxies with $\mffeat \geq 0.4$ also have $z > 1.5$, whereas only 5 per cent of galaxies with $\mffeat \geq 0.7$ have $z > 1.5$. This effect is also discussed in \citet{bamford09}, and we discuss image resolution in more detail in Section \ref{sec:resolution}.

For smooth galaxies, as expected, the effect on redshift and brightness distributions is much smaller, with the peak of the absolute $V$ magnitude distribution of the ``clean'' smooth sample ($\mfsmooth \geq 0.8$) at $M_{V,~\rm peak} = -20.4$ and galaxies across the complete range of both variables included in the ``clean'' smooth sample (Figure \ref{fig:Mz_thresh}). 

As an example of the use of these morphologies, Figure \ref{fig:UVJ} shows rest-frame $UVJ$ colour-colour plots for different morphological samples at different redshifts. Within each redshift range, the smooth sample is chosen to match the rest-$V$ luminosity range of the spiral or clumpy sample being compared. The clean spiral sample has colours generally associated with star-forming galaxies with redder colours extending into the dusty region of $UVJ$ space, but avoiding the passive region \citep[e.g.,][]{williams09,muzzin13a}. Clumpy galaxies tend to be blue at all redshifts examined. Smooth galaxies generally span the range of observed colours compared to the full sample. This result is similar to that seen by \citet{kartaltepe15}, although that study measures somewhat different quantities in their visual classifications. We compare our classifications to those of \citeauthor{kartaltepe15} in detail in Section \ref{sec:comparison}, and we discuss the smooth galaxy sample in further detail in Section \ref{sec:result}.

%%%%% [FIGURE: Mag vs z for different thresholds (smooth and featured)] %%%%%
\begin{figure*}
\includegraphics[scale=0.85]{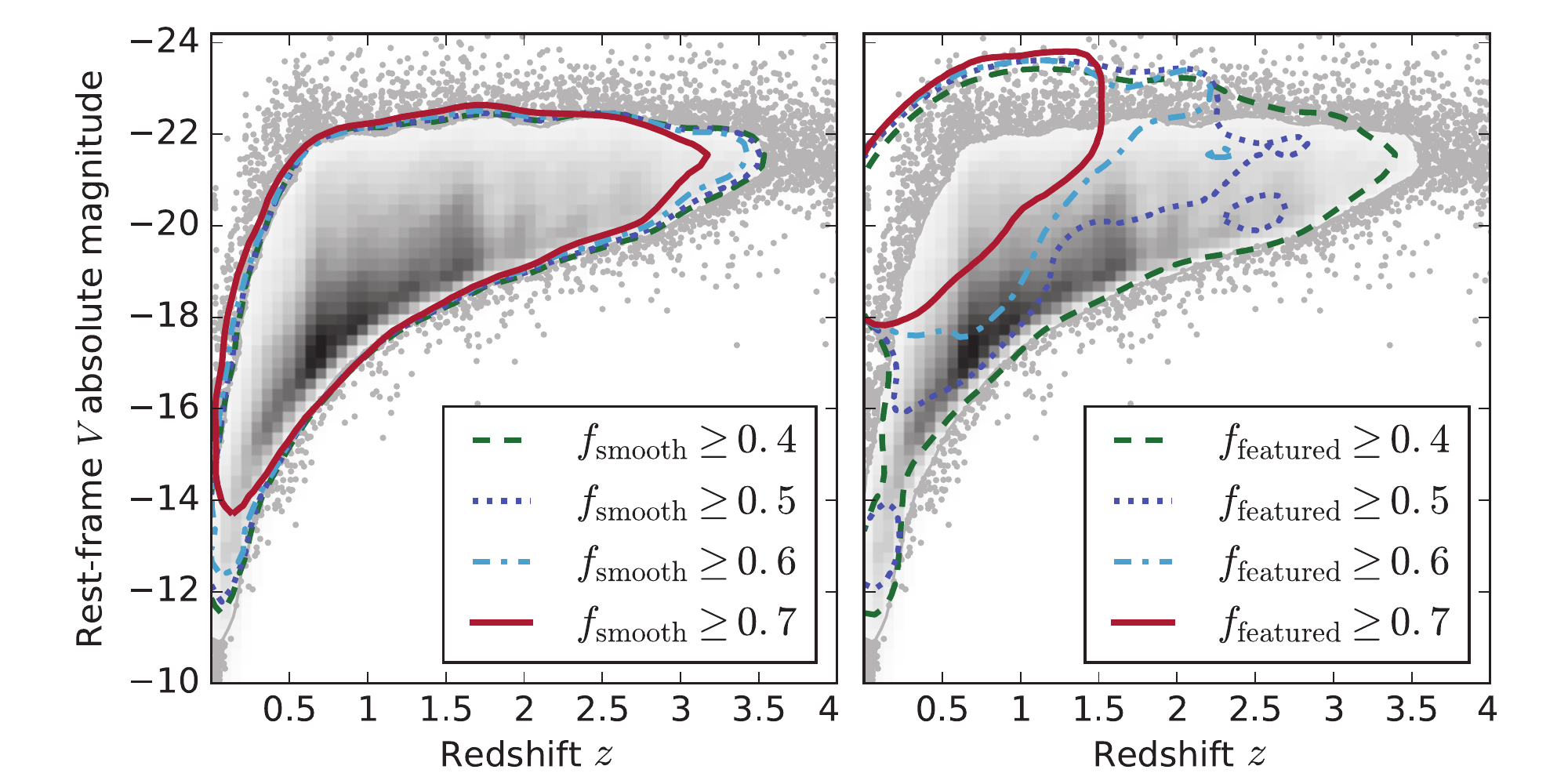}
\caption{
Both panels: absolute rest-frame $V$-band magnitude versus redshift, with grayscale 2-D density plots showing the full sample of galaxies (Gaussian KDE method; gray points show galaxies outside the region including 98 per cent of the distribution). The left panel also shows sub-samples of smooth galaxies selected using different \fsmooth\ thresholds; the right panel shows sub-samples of featured galaxies selected using different \ffeat\ thresholds. Each contour shows the region enclosing 98 per cent of galaxies selecting using each threshold (green dashed for $f \geq 0.4$, dark blue dotted for $f \geq 0.5$, light blue dot-dashed for $f \geq 0.6$, and red solid for $f \geq 0.7$, in each panel). The region of $V-z$ parameter space sampled by a smooth galaxy sample is relatively insensitive to the choice of threshold, whereas more restrictive cuts on \ffeat\ in selecting a featured sample select more restricted distributions of luminosity and redshift. 
}
\label{fig:Mz_thresh}
\end{figure*}
%%%%% END FIGURE %%%%%

%%%%% [FIGURE: UVJ for all, smooth, clumpy, spiral samples] %%%%%
\begin{figure*}
\includegraphics[scale=0.625]{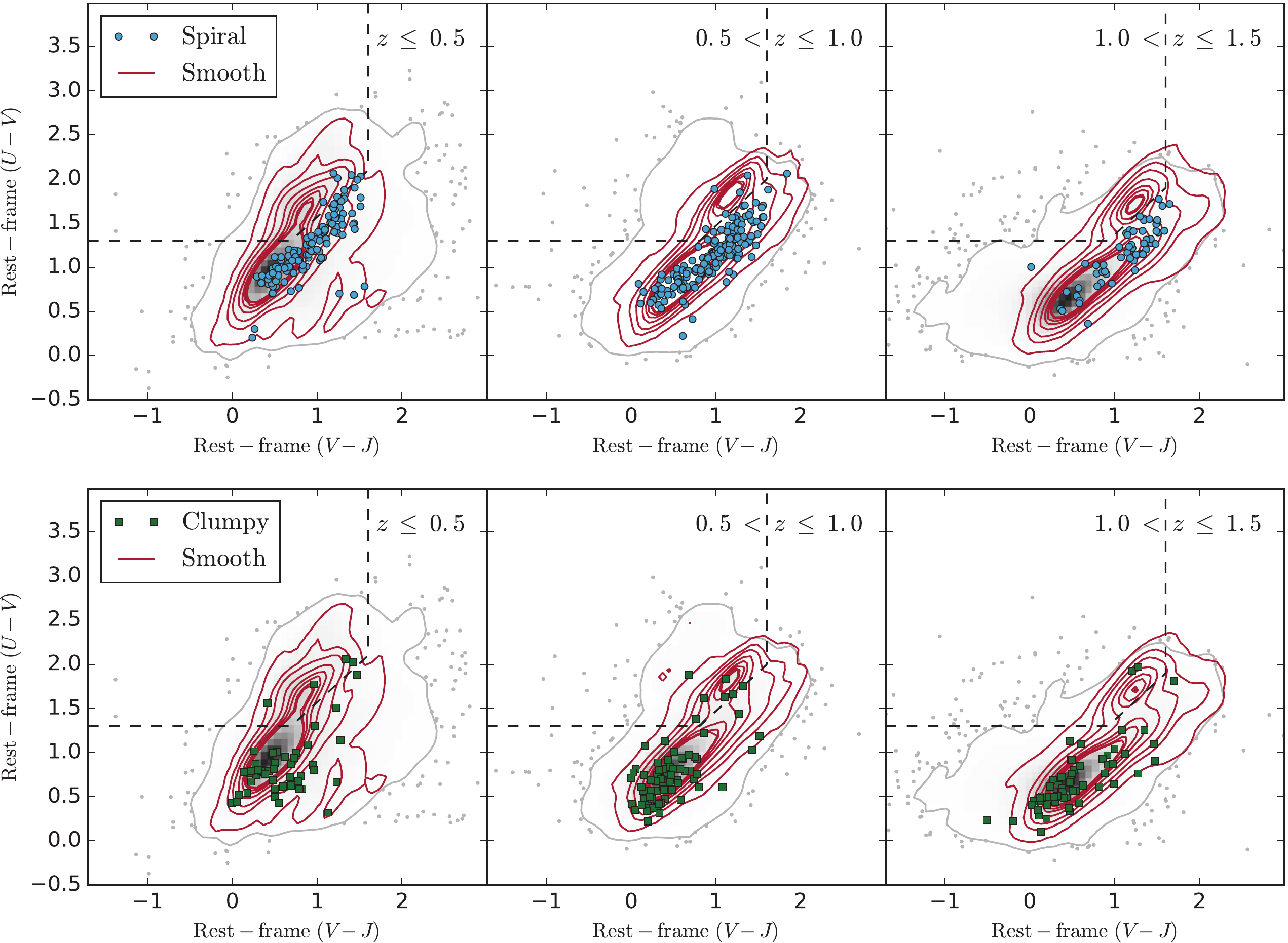}
\caption{
Rest-frame $U-V$ vs $V-J$ colour-colour plots for different classification selections. From left to right, each column shows sources in increasing redshift bins, with $z \leq 0.5$ in the leftmost column, $0.5 < z \leq 1$ in the centre, and $1 < z \leq 1.5$ at right. In all panels, grayscale points, contours and shading show the full sample within that redshift range, and red lines show contours representing the clean ``smooth" sample.  In the top row, blue circles show galaxies in the clean ``spiral" sample. In the bottom row, green squares show galaxies in the clean ``clumpy" sample. (See Table \ref{table:clean} for a definition of each sample.) In each panel, smooth and spiral (or smooth and clumpy) samples are restricted to the same rest-frame $V$-band luminosity range. Dashed lines show the empirical selection criteria for quiescent galaxies in each redshift bin \citep[from][]{williams09}. Each morphologically distinct sample occupies a different part of colour-colour space. 
}
\label{fig:UVJ}
\end{figure*}
%%%%% END FIGURE %%%%%

\begin{table*}
\resizebox{\textwidth}{!}{%
\begin{tabular}{@{}lccccccc}
\hline
\multicolumn{1}{l}{Data Column Name} &
\multicolumn{6}{c}{Subjects} &
\multicolumn{1}{c}{}
\\
\hline
\hline
ID & {\small COS\_11292} & COS\_26682 & GDS\_17388 & GDS\_25569 & UDS\_14391 & UDS\_23274 & $\ldots$\\
RA & 150.110092 & 150.131912 & 53.126442 & 53.189608 & 34.323514 & 34.490365 & \\
Dec & 2.314499 & 2.516023 & -27.756544 & -27.695279 & -5.198308 & -5.143040 & \\
num\_classifications & 38 & 76 & 74 & 39 & 73 & 78 &\\
num\_classifications\_weighted & 35.04 & 67.15 & 82.23 & 36.62 & 62.00 & 72.76 &\\
t00\_smooth\_or\_featured\_a0\_smooth\_frac & 0.61 & 0.30 & 0.19 & 0.44 & 0.26 & 0.44 &\\
t00\_smooth\_or\_featured\_a1\_features\_frac & 0.03 & 0.25 & 0.62 & 0.18 & 0.23 & 0.42 &\\
t00\_smooth\_or\_featured\_a2\_artifact\_frac & 0.37 & 0.45 & 0.19 & 0.38 & 0.51 & 0.14 &\\
t00\_smooth\_or\_featured\_a0\_smooth\_weighted\_frac & 0.84 & 0.41 & 0.24 & 0.63 & 0.39 & 0.51 &\\
t00\_smooth\_or\_featured\_a1\_features\_weighted\_frac & 0.04 & 0.31 & 0.74 & 0.24 & 0.30 & 0.46 &\\
t00\_smooth\_or\_featured\_a2\_artifact\_weighted\_frac & 0.11 & 0.28 & 0.02 & 0.12 & 0.31 & 0.03 &\\
t00\_smooth\_or\_featured\_count & 38 & 76 & 74 & 39 & 73 & 78 &\\
t00\_smooth\_or\_featured\_weight & 42.13 & 88.04 & 82.23 & 42.47 & 75.17 & 87.94 &\\
t01\_how\_rounded\_a0\_completely\_frac & 0.87 & 0.17 & 0.86 & 0.59 & 0.16 & 0.21 &\\
t01\_how\_rounded\_a1\_inbetween\_frac & 0.13 & 0.83 & 0.14 & 0.41 & 0.37 & 0.74 &\\
t01\_how\_rounded\_a2\_cigarshaped\_frac & 0.00 & 0.00 & 0.00 & 0.00 & 0.47 & 0.06 &\\
t01\_how\_rounded\_a0\_completely\_weighted\_frac & 0.87 & 0.17 & 0.85 & 0.58 & 0.16 & 0.21 &\\
t01\_how\_rounded\_a1\_inbetween\_weighted\_frac & 0.13 & 0.83 & 0.15 & 0.42 & 0.37 & 0.74 &\\
t01\_how\_rounded\_a2\_cigarshaped\_weighted\_frac & 0.00 & 0.00 & 0.00 & 0.00 & 0.47 & 0.05 &\\
t01\_how\_rounded\_count & 23 & 23 & 14 & 17 & 19 & 34 &\\
t01\_how\_rounded\_weight & 23.00 & 22.99 & 13.78 & 16.55 & 18.96 & 32.68 &\\
t02\_clumpy\_appearance\_a0\_yes\_frac & 0.00 & 0.74 & 0.13 & 0.86 & 0.41 & 0.45 &\\
t02\_clumpy\_appearance\_a1\_no\_frac & 1.00 & 0.26 & 0.87 & 0.14 & 0.59 & 0.55 &\\
t02\_clumpy\_appearance\_a0\_yes\_weighted\_frac & 0.00 & 0.74 & 0.12 & 0.86 & 0.39 & 0.45 &\\
t02\_clumpy\_appearance\_a1\_no\_weighted\_frac & 1.00 & 0.26 & 0.88 & 0.14 & 0.61 & 0.55 &\\
t02\_clumpy\_appearance\_count & 1 & 19 & 46 & 7 & 17 & 33 &\\
t02\_clumpy\_appearance\_weight & 1.00 & 19.00 & 45.71 & 7.00 & 16.28 & 32.46 &\\
t03\_how\_many\_clumps\_a0\_1\_frac & 0.00 & 0.00 & 0.50 & 0.17 & 0.00 & 0.60 &\\
t03\_how\_many\_clumps\_a1\_2\_frac & 0.00 & 0.36 & 0.00 & 0.33 & 0.00 & 0.00 &\\
t03\_how\_many\_clumps\_a2\_3\_frac & 0.00 & 0.29 & 0.13 & 0.00 & 0.29 & 0.00 &\\
t03\_how\_many\_clumps\_a3\_4\_frac & 0.00 & 0.00 & 0.00 & 0.00 & 0.14 & 0.13 &\\
t03\_how\_many\_clumps\_a4\_5\_plus\_frac & 0.00 & 0.00 & 0.38 & 0.00 & 0.43 & 0.00 &\\
t03\_how\_many\_clumps\_a5\_cant\_tell\_frac & 0.00 & 0.36 & 0.00 & 0.50 & 0.14 & 0.27 &\\
t03\_how\_many\_clumps\_a0\_1\_weighted\_frac & 0.00 & 0.00 & 0.50 & 0.17 & 0.00 & 0.62 &\\
t03\_how\_many\_clumps\_a1\_2\_weighted\_frac & 0.00 & 0.36 & 0.00 & 0.33 & 0.00 & 0.00 &\\
t03\_how\_many\_clumps\_a2\_3\_weighted\_frac & 0.00 & 0.29 & 0.15 & 0.00 & 0.32 & 0.00 &\\
t03\_how\_many\_clumps\_a3\_4\_weighted\_frac & 0.00 & 0.00 & 0.00 & 0.00 & 0.16 & 0.11 &\\
t03\_how\_many\_clumps\_a4\_5\_plus\_weighted\_frac & 0.00 & 0.00 & 0.35 & 0.00 & 0.36 & 0.00 &\\
t03\_how\_many\_clumps\_a5\_cant\_tell\_weighted\_frac & 0.00 & 0.36 & 0.00 & 0.50 & 0.16 & 0.27 &\\
t03\_how\_many\_clumps\_count & 0 & 14 & 6 & 6 & 7 & 15 &\\
t03\_how\_many\_clumps\_weight & 0.00 & 14.00 & 5.36 & 6.00 & 6.28 & 14.46 &\\
t04\_clump\_configuration\_a0\_straight\_line\_frac & 0.00 & 0.00 & 0.00 & 0.00 & 0.57 & 0.17 &\\
t04\_clump\_configuration\_a1\_chain\_frac & 0.00 & 0.00 & 0.00 & 0.00 & 0.00 & 0.17 &\\
t04\_clump\_configuration\_a2\_cluster\_or\_irregular\_frac & 0.00 & 1.00 & 1.00 & 1.00 & 0.43 & 0.67 &\\
t04\_clump\_configuration\_a3\_spiral\_frac & 0.00 & 0.00 & 0.00 & 0.00 & 0.00 & 0.00 &\\
t04\_clump\_configuration\_a0\_straight\_line\_weighted\_frac & 0.00 & 0.00 & 0.00 & 0.00 & 0.64 & 0.18 &\\
t04\_clump\_configuration\_a1\_chain\_weighted\_frac & 0.00 & 0.00 & 0.00 & 0.00 & 0.00 & 0.18 &\\
t04\_clump\_configuration\_a2\_cluster\_or\_irregular\_weighted\_frac & 0.00 & 1.00 & 1.00 & 1.00 & 0.36 & 0.63 &\\
t04\_clump\_configuration\_a3\_spiral\_weighted\_frac & 0.00 & 0.00 & 0.00 & 0.00 & 0.00 & 0.00 &\\
t04\_clump\_configuration\_count & 0 & 9 & 4 & 3 & 7 & 6 &\\
t04\_clump\_configuration\_weight & 0.00 & 9.00 & 3.36 & 3.00 & 6.28 & 5.46 &\\
t05\_is\_one\_clump\_brightest\_a0\_yes\_frac & 0.00 & 0.07 & 0.25 & 1.00 & 0.43 & 0.67 &\\
t05\_is\_one\_clump\_brightest\_a1\_no\_frac & 0.00 & 0.93 & 0.75 & 0.00 & 0.57 & 0.33 &\\
t05\_is\_one\_clump\_brightest\_a0\_yes\_weighted\_frac & 0.00 & 0.07 & 0.30 & 1.00 & 0.48 & 0.63 &\\
t05\_is\_one\_clump\_brightest\_a1\_no\_weighted\_frac & 0.00 & 0.93 & 0.70 & 0.00 & 0.52 & 0.37 &\\
t05\_is\_one\_clump\_brightest\_count & 0 & 14 & 4 & 5 & 7 & 6 &\\
t05\_is\_one\_clump\_brightest\_weight & 0.00 & 14.00 & 3.36 & 5.00 & 6.28 & 5.46 &\\
t06\_brightest\_clump\_central\_a0\_yes\_frac & 0.00 & 0.00 & 1.00 & 0.60 & 1.00 & 0.75 &\\
t06\_brightest\_clump\_central\_a1\_no\_frac & 0.00 & 1.00 & 0.00 & 0.40 & 0.00 & 0.25 &\\
t06\_brightest\_clump\_central\_a0\_yes\_weighted\_frac & 0.00 & 0.00 & 1.00 & 0.60 & 1.00 & 0.83 &\\
t06\_brightest\_clump\_central\_a1\_no\_weighted\_frac & 0.00 & 1.00 & 0.00 & 0.40 & 0.00 & 0.17 &\\
t06\_brightest\_clump\_central\_count & 0 & 1 & 1 & 5 & 3 & 4 &\\
t06\_brightest\_clump\_central\_weight & 0.00 & 1.00 & 1.00 & 5.00 & 3.00 & 3.46 &\\
t07\_galaxy\_symmetrical\_a0\_yes\_frac & 0.00 & 0.08 & 0.67 & 0.25 & 0.29 & 0.21 &\\
t07\_galaxy\_symmetrical\_a1\_no\_frac & 0.00 & 0.92 & 0.33 & 0.75 & 0.71 & 0.79 &\\
t07\_galaxy\_symmetrical\_a0\_yes\_weighted\_frac & 0.00 & 0.08 & 0.70 & 0.25 & 0.32 & 0.22 &\\
t07\_galaxy\_symmetrical\_a1\_no\_weighted\_frac & 0.00 & 0.92 & 0.30 & 0.75 & 0.68 & 0.78 &\\
t07\_galaxy\_symmetrical\_count & 0 & 13 & 6 & 4 & 7 & 14 &\\
t07\_galaxy\_symmetrical\_weight & 0.00 & 13.00 & 5.36 & 4.00 & 6.28 & 13.87 &\\
$\ldots$    \\
\hline
\end{tabular}}
\caption{Raw and weighted classifications for the full Galaxy Zoo CANDELS sample. The complete version of this table is available in electronic form and at http://data.galaxyzoo.org. The printed table shows a transposed subset of the full table to illustrate its format and content. The complete version includes raw and weighted morphological classifications for all tasks in the classification tree (Figure \ref{fig:tree} and Table \ref{table:tree}) for each of the 49,555 unique sources in the sample, as well as flags {\tt clean\_smooth, clean\_featured, clean\_clumpy, clean\_edge\_on, clean\_spiral} which mark galaxies in the ``clean'' subsamples described in Section \ref{sec:release} and Table \ref{table:clean}, and the flag {\tt smooth\_disk} which marks smooth galaxies with significant disk components, as described in Section \ref{sec:result}. }
\label{table:data-main}
\end{table*}

\begin{table*}
\resizebox{\textwidth}{!}{%
\begin{tabular}{@{}lccccccc}
\hline
\multicolumn{1}{l}{Data Column Name} &
\multicolumn{6}{c}{Subjects} &
\multicolumn{1}{c}{}
\\
\hline
\hline
ID & {\small COS\_11292} & COS\_26682 & GDS\_17388 & GDS\_25569 & UDS\_14391 & UDS\_23274 & $\ldots$\\
$\ldots$     \\
t08\_clumps\_embedded\_larger\_object\_a0\_yes\_frac & 0.00 & 0.31 & 0.83 & 0.25 & 0.86 & 0.57 &\\
t08\_clumps\_embedded\_larger\_object\_a1\_no\_frac & 0.00 & 0.69 & 0.17 & 0.75 & 0.14 & 0.43 &\\
t08\_clumps\_embedded\_larger\_object\_a0\_yes\_weighted\_frac & 0.00 & 0.31 & 0.81 & 0.25 & 0.96 & 0.57 &\\
t08\_clumps\_embedded\_larger\_object\_a1\_no\_weighted\_frac & 0.00 & 0.69 & 0.19 & 0.75 & 0.04 & 0.43 &\\
t08\_clumps\_embedded\_larger\_object\_count & 0 & 13 & 6 & 4 & 7 & 14 &\\
t08\_clumps\_embedded\_larger\_object\_weight & 0.00 & 13.00 & 5.36 & 4.00 & 6.28 & 13.87 &\\
t09\_disk\_edge\_on\_a0\_yes\_frac & 0.00 & 0.20 & 0.00 & 0.00 & 0.60 & 0.11 &\\
t09\_disk\_edge\_on\_a1\_no\_frac & 1.00 & 0.80 & 1.00 & 1.00 & 0.40 & 0.89 &\\
t09\_disk\_edge\_on\_a0\_yes\_weighted\_frac & 0.00 & 0.20 & 0.00 & 0.00 & 0.60 & 0.11 &\\
t09\_disk\_edge\_on\_a1\_no\_weighted\_frac & 1.00 & 0.80 & 1.00 & 1.00 & 0.40 & 0.89 &\\
t09\_disk\_edge\_on\_count & 1 & 5 & 40 & 1 & 10 & 18 &\\
t09\_disk\_edge\_on\_weight & 1.00 & 5.00 & 40.35 & 1.00 & 10.00 & 18.00 &\\
t10\_edge\_on\_bulge\_a0\_yes\_frac & 0.00 & 0.00 & 0.00 & 0.00 & 0.17 & 1.00 &\\
t10\_edge\_on\_bulge\_a1\_no\_frac & 0.00 & 1.00 & 0.00 & 0.00 & 0.83 & 0.00 &\\
t10\_edge\_on\_bulge\_a0\_yes\_weighted\_frac & 0.00 & 0.00 & 0.00 & 0.00 & 0.17 & 1.00 &\\
t10\_edge\_on\_bulge\_a1\_no\_weighted\_frac & 0.00 & 1.00 & 0.00 & 0.00 & 0.83 & 0.00 &\\
t10\_edge\_on\_bulge\_count & 0 & 1 & 0 & 0 & 6 & 2 &\\
t10\_edge\_on\_bulge\_weight & 0.00 & 1.00 & 0.00 & 0.00 & 6.00 & 2.00 &\\
t11\_bar\_feature\_a0\_yes\_frac & 0.00 & 0.00 & 0.05 & 0.00 & 0.50 & 0.00 &\\
t11\_bar\_feature\_a1\_no\_frac & 1.00 & 1.00 & 0.95 & 1.00 & 0.50 & 1.00 &\\
t11\_bar\_feature\_a0\_yes\_weighted\_frac & 0.00 & 0.00 & 0.05 & 0.00 & 0.50 & 0.00 &\\
t11\_bar\_feature\_a1\_no\_weighted\_frac & 1.00 & 1.00 & 0.95 & 1.00 & 0.50 & 1.00 &\\
t11\_bar\_feature\_count & 1 & 4 & 40 & 1 & 4 & 16 &\\
t11\_bar\_feature\_weight & 1.00 & 4.00 & 40.35 & 1.00 & 4.00 & 16.00 &\\
t12\_spiral\_pattern\_a0\_yes\_frac & 0.00 & 0.50 & 0.60 & 0.00 & 0.25 & 0.69 &\\
t12\_spiral\_pattern\_a1\_no\_frac & 1.00 & 0.50 & 0.40 & 1.00 & 0.75 & 0.31 &\\
t12\_spiral\_pattern\_a0\_yes\_weighted\_frac & 0.00 & 0.50 & 0.59 & 0.00 & 0.25 & 0.69 &\\
t12\_spiral\_pattern\_a1\_no\_weighted\_frac & 1.00 & 0.50 & 0.41 & 1.00 & 0.75 & 0.31 &\\
t12\_spiral\_pattern\_count & 1 & 4 & 40 & 1 & 4 & 16 &\\
t12\_spiral\_pattern\_weight & 1.00 & 4.00 & 40.35 & 1.00 & 4.00 & 16.00 &\\
t13\_spiral\_arm\_winding\_a0\_tight\_frac & 0.00 & 0.00 & 0.79 & 0.00 & 0.00 & 0.36 &\\
t13\_spiral\_arm\_winding\_a1\_medium\_frac & 0.00 & 0.50 & 0.17 & 0.00 & 0.00 & 0.45 &\\
t13\_spiral\_arm\_winding\_a2\_loose\_frac & 0.00 & 0.50 & 0.04 & 0.00 & 1.00 & 0.18 &\\
t13\_spiral\_arm\_winding\_a0\_tight\_weighted\_frac & 0.00 & 0.00 & 0.79 & 0.00 & 0.00 & 0.36 &\\
t13\_spiral\_arm\_winding\_a1\_medium\_weighted\_frac & 0.00 & 0.50 & 0.17 & 0.00 & 0.00 & 0.45 &\\
t13\_spiral\_arm\_winding\_a2\_loose\_weighted\_frac & 0.00 & 0.50 & 0.04 & 0.00 & 1.00 & 0.18 &\\
t13\_spiral\_arm\_winding\_count & 0 & 2 & 24 & 0 & 1 & 11 &\\
t13\_spiral\_arm\_winding\_weight & 0.00 & 2.00 & 24.00 & 0.00 & 1.00 & 11.00 &\\
t14\_spiral\_arm\_count\_a0\_1\_frac & 0.00 & 0.00 & 0.08 & 0.00 & 0.00 & 0.27 &\\
t14\_spiral\_arm\_count\_a1\_2\_frac & 0.00 & 1.00 & 0.38 & 0.00 & 0.00 & 0.18 &\\
t14\_spiral\_arm\_count\_a2\_3\_frac & 0.00 & 0.00 & 0.08 & 0.00 & 0.00 & 0.00 &\\
t14\_spiral\_arm\_count\_a3\_4\_frac & 0.00 & 0.00 & 0.00 & 0.00 & 1.00 & 0.00 &\\
t14\_spiral\_arm\_count\_a4\_5\_plus\_frac & 0.00 & 0.00 & 0.00 & 0.00 & 0.00 & 0.00 &\\
t14\_spiral\_arm\_count\_a5\_cant\_tell\_frac & 0.00 & 0.00 & 0.46 & 0.00 & 0.00 & 0.55 &\\
t14\_spiral\_arm\_count\_a0\_1\_weighted\_frac & 0.00 & 0.00 & 0.08 & 0.00 & 0.00 & 0.27 &\\
t14\_spiral\_arm\_count\_a1\_2\_weighted\_frac & 0.00 & 1.00 & 0.38 & 0.00 & 0.00 & 0.18 &\\
t14\_spiral\_arm\_count\_a2\_3\_weighted\_frac & 0.00 & 0.00 & 0.08 & 0.00 & 0.00 & 0.00 &\\
t14\_spiral\_arm\_count\_a3\_4\_weighted\_frac & 0.00 & 0.00 & 0.00 & 0.00 & 1.00 & 0.00 &\\
t14\_spiral\_arm\_count\_a4\_5\_plus\_weighted\_frac & 0.00 & 0.00 & 0.00 & 0.00 & 0.00 & 0.00 &\\
t14\_spiral\_arm\_count\_a5\_cant\_tell\_weighted\_frac & 0.00 & 0.00 & 0.46 & 0.00 & 0.00 & 0.55 &\\
t14\_spiral\_arm\_count\_count & 0 & 2 & 24 & 0 & 1 & 11 &\\
t14\_spiral\_arm\_count\_weight & 0.00 & 2.00 & 24.00 & 0.00 & 1.00 & 11.00 &\\
t15\_bulge\_prominence\_a0\_no\_bulge\_frac & 1.00 & 0.75 & 0.03 & 0.00 & 0.25 & 0.25 &\\
t15\_bulge\_prominence\_a1\_obvious\_frac & 0.00 & 0.25 & 0.90 & 0.00 & 0.25 & 0.69 &\\
t15\_bulge\_prominence\_a2\_dominant\_frac & 0.00 & 0.00 & 0.08 & 1.00 & 0.50 & 0.06 &\\
t15\_bulge\_prominence\_a0\_no\_bulge\_weighted\_frac & 1.00 & 0.75 & 0.02 & 0.00 & 0.25 & 0.25 &\\
t15\_bulge\_prominence\_a1\_obvious\_weighted\_frac & 0.00 & 0.25 & 0.90 & 0.00 & 0.25 & 0.69 &\\
t15\_bulge\_prominence\_a2\_dominant\_weighted\_frac & 0.00 & 0.00 & 0.07 & 1.00 & 0.50 & 0.06 &\\
t15\_bulge\_prominence\_count & 1 & 4 & 40 & 1 & 4 & 16 &\\
t15\_bulge\_prominence\_weight & 1.00 & 4.00 & 40.35 & 1.00 & 4.00 & 16.00 &\\
t16\_merging\_tidal\_debris\_a0\_merging\_frac & 0.21 & 0.40 & 0.45 & 0.04 & 0.22 & 0.07 &\\
t16\_merging\_tidal\_debris\_a1\_tidal\_debris\_frac & 0.00 & 0.07 & 0.02 & 0.42 & 0.11 & 0.16 &\\
t16\_merging\_tidal\_debris\_a2\_both\_frac & 0.00 & 0.29 & 0.13 & 0.04 & 0.11 & 0.01 &\\
t16\_merging\_tidal\_debris\_a3\_neither\_frac & 0.79 & 0.24 & 0.40 & 0.50 & 0.56 & 0.75 &\\
t16\_merging\_tidal\_debris\_a0\_merging\_weighted\_frac & 0.21 & 0.40 & 0.45 & 0.04 & 0.21 & 0.08 &\\
t16\_merging\_tidal\_debris\_a1\_tidal\_debris\_weighted\_frac & 0.00 & 0.07 & 0.02 & 0.42 & 0.11 & 0.17 &\\
t16\_merging\_tidal\_debris\_a2\_both\_weighted\_frac & 0.00 & 0.29 & 0.12 & 0.04 & 0.11 & 0.01 &\\
t16\_merging\_tidal\_debris\_a3\_neither\_weighted\_frac & 0.79 & 0.24 & 0.41 & 0.49 & 0.57 & 0.75 &\\
t16\_merging\_tidal\_debris\_count & 24 & 42 & 60 & 24 & 36 & 67 &\\
t16\_merging\_tidal\_debris\_weight & 24.00 & 41.99 & 59.49 & 23.55 & 35.24 & 65.14 &\\
\hline
\end{tabular}}
\contcaption{}
%\label{table:data-main}
\end{table*}

\begin{table}
 \begin{tabular}{@{}lllr}
 \hline
\multicolumn{1}{l}{Clean sample} &
\multicolumn{1}{c}{Tasks} &
\multicolumn{1}{c}{Selection} &
\multicolumn{1}{c}{$N_{\rm sample}$}
\\ 
\hline
\hline
\textbf{Smooth} & T00 & $f_{\rm smooth} \geq 0.8$ & 6770\\ \hline
\textbf{Featured} & T00 & $\mffeat \geq 0.7$ & 312\\ \hline
\textbf{Clumpy} & \parbox[t]{0.75cm}{T00\\T02\\T16} & \parbox[t]{3.5cm}{$\mffeat \geq 0.4$\\$f_{\rm clumpy} \geq 0.7$, $N_{\rm T02} \geq 10$\\$f_{\rm neither} \geq 0.25$} & 333\\ \hline
\textbf{Edge-on} & \parbox[t]{0.75cm}{T00\\T02\\T09} & \parbox[t]{3.5cm}{$\mffeat \geq 0.4$\\$f_{\rm not\ clumpy} \geq 0.3$\\$f_{\rm edge-on} > 0.7$, $N_{\rm T09} \geq 10$} & 223\\ \hline
\textbf{Spiral}  & \parbox[t]{0.75cm}{T00\\T02\\T09\\T12} & \parbox[t]{3.5cm}{$\mffeat \geq 0.4$\\$f_{\rm not\ clumpy} \geq 0.3$\\$f_{\rm not\ edge-on} \geq 0.5$\\$f_{\rm spiral} \geq 0.8$, $N_{\rm T12} \geq 10$} & 383\\ \hline
 \end{tabular}
 \caption{Clean samples, designated in the Galaxy Zoo CANDELS weighted classification catalog with flags. Each sample is selected using criteria from at least 1 task in the decision tree (Figure \ref{fig:tree} and Table \ref{table:tree}); samples selecting for features characterised in lower branches of the tree include selection criteria on dependent tasks. The selection for the Clumpy clean sample also includes a rejection of subjects having a substantial fraction of classifications for ``merging'', ``tidal debris'', or ``both''. The Clean samples are relatively free from contaminants, and are correspondingly incomplete.
\label{table:clean}}
\end{table}

\subsection{Depth Corrections}\label{sec:depth}

%%%%% [FIGURE: Depth Corrections] %%%%%
\begin{figure*}
\includegraphics[scale=0.5]{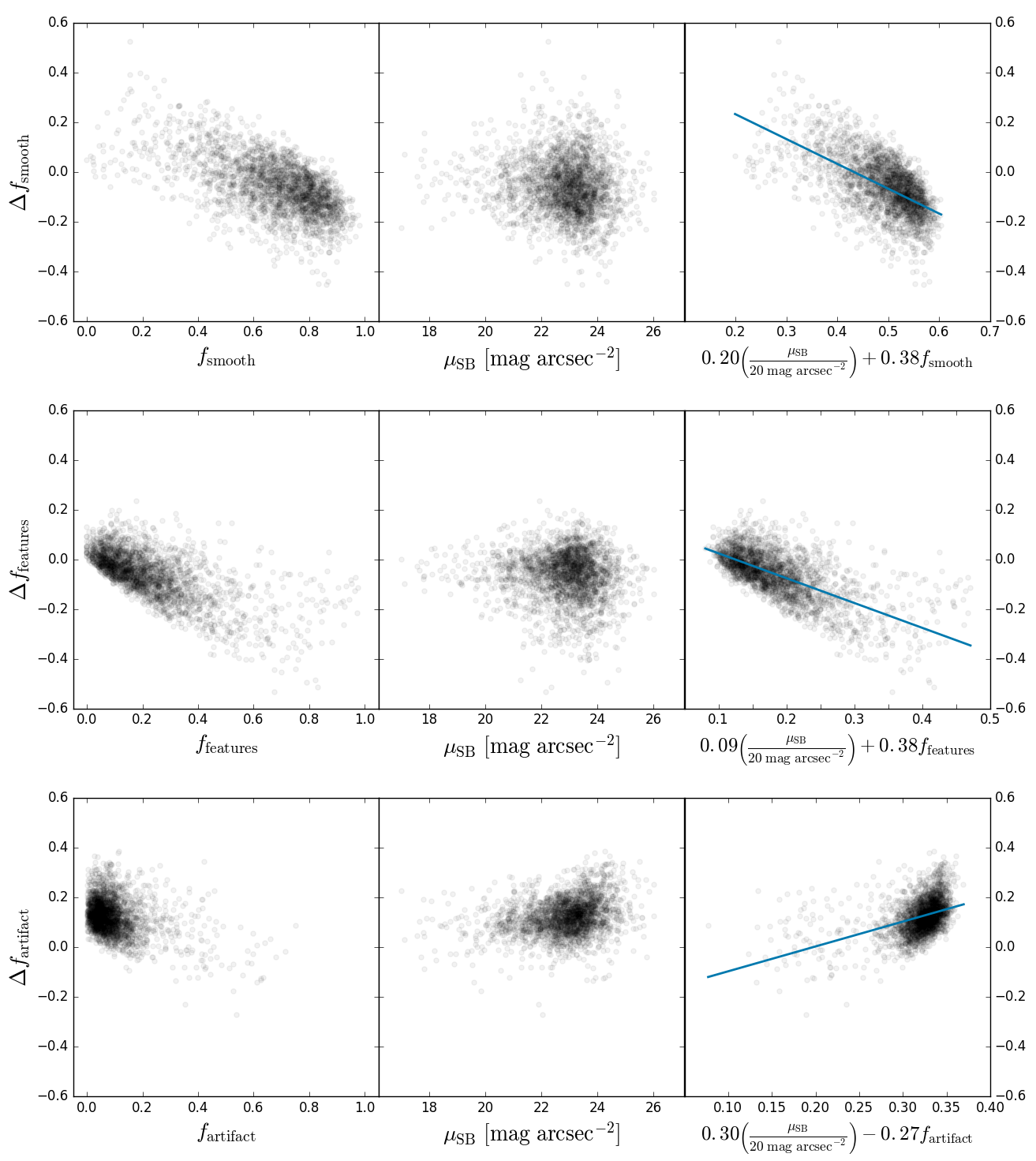}
\caption{
Depth corrections to classifications for individual responses to task T00: ``smooth'' (top row), ``features or disk'' (middle row), and ``star or artifact'' (bottom row). The change in classification ($\Delta f$) between the deep- and wide-field-depth observations of the same subjects is fit as a function of deep-field-depth morphology (vote fraction $f$, left column) and galaxy characteristic surface brightness (\musb , middle column). Each point is a subject which received at least 10 responses within the task. The best-fit plane is shown in projection (blue line) in the right-hand column. 
%Fits were performed only for tasks where at least 75 subjects had 10 or more responses to the task; otherwise the correction is assumed to be 0. 
}
\label{fig:depthcorr}
\end{figure*}
%%%%% END FIGURE %%%%%

As outlined in Section \ref{sec:images}, the depth of the CANDELS survey varies substantially between two types of fields, from shallower ``wide'' fields with $\sim 1$ orbit depth in $F160W$ to ``deep'' fields with $\gtrsim 4$ times the overall depth of the wide fields. 

Because different morphological features have different characteristic surface brightnesses and light profiles (e.g. clumps, tidal signatures, spheroids, disks), we expect the morphological classifications of subjects to vary somewhat based on the imaging depth. For a survey such as CANDELS that is already relatively deep, we expect this effect to generally be small, but nevertheless using classifications based on imaging from both the wide and deep fields could complicate some scientific inquiries.

To measure and correct for this, shallower (2-epoch) images of a sub-sample of ``deep'' subjects were created and added to the active subject sets. These images (from 2,518 subjects in the GOODS-South deep field; hereafter the ``measured-correction'' subset) are of comparable depth to the wide fields. Below we describe how comparison of the shallower and deeper weighted consensus classifications of these subjects allows us to determine typical depth corrections to all deep-field subject morphologies as a function of deep-exposure morphology and galaxy surface brightness.

We define the observed surface brightness of a galaxy using the magnitude and size reported in the CANDELS photometric catalogs for each field. Specifically, we use the $F160W$ \textsc{auto} fluxes and the radius containing 80\% of the galaxy light, $r_{80}$, to determine a representative surface brightness for each galaxy. We also use the measured axis ratios $b/a$, dividing the flux by the area contained within an ellipse of area $\pi a_{80} b_{80}$, in square arcseconds. We then convert to magnitudes, resulting in a single surface brightness \musb\ in mag/arcsec$^2$ for each subject.

Figure \ref{fig:depthcorr} shows the difference between shallower and deeper weighted consensus classifications as a function of surface brightness and deep-exposure morphology for the initial task in the classification tree, T00. When determining depth corrections for each task, we consider subjects which received more than 10 answers to the question presented by the task, and we also remove 42 bright stars with $\mmusb > 17$ (the mean weighted ``star or artifact'' vote fraction for these is $f_{\rm artifact} = 0.89$). For tasks T01-T16, we additionally remove subjects with $f_{\rm artifact} \geq 0.5$. We then determine a best-fit plane to the change in vote fraction, $\Delta f_r \equiv f_{\rm shallow} - f_{\rm deep}$, for each response as a function of galaxy surface brightness and the vote fraction $f_{\rm deep}$ for the deep-exposure image. 
% Turns out this wasn't the case for any of the tasks in the end
%If fewer than 75 subjects received at least 10 responses to a particular question, the correction is assumed to be 0 for that task.

In general, the correction to the vote fraction is a stronger function of the vote fraction than of the surface brightness, though most corrections do depend on both. For example, the correction $\Delta \mffeat$ (middle row of Figure \ref{fig:depthcorr}) is clearly a linear function of the deep-exposure vote fraction \ffeat , such that highly ``featured'' galaxies tend to have lower featured vote fractions in the shallower images (as expected), with a slight turnover at $\mffeat > 0.9$ indicating that such galaxies, with obvious features, are still identifiable as featured even at shallower depth. The $\Delta \mffeat-\mmusb$ relation is nearly flat, but the scatter is higher for lower surface-brightness galaxies. As expected, for very bright galaxies the change in depth makes little difference to the classification, whereas for fainter galaxies the change depends more on details of features that vary from galaxy to galaxy. It is therefore important to note that the best-fit corrections (right column of Figure \ref{fig:depthcorr}) are average values across the whole sample and can be highly uncertain for an individual galaxy.

The best-fit planes for each response to each task can be used to predict the wide-field depth classifications for galaxies in the deep fields, both in this paper and in future releases of Galaxy Zoo CANDELS data.  For the 8,130 subjects with $F160W$ limiting magnitudes at least as faint as the brightest limiting magnitude in the measured-correction subset ($F160W = 28.26$) but for which we do not also have separate wide-field depth classifications, we use their vote fractions for each task and response, $f_r$, and their surface brightnesses \musb , to interpolate corrections $\Delta f_r$ along the best-fit plane for each task and response. Where a $\left( f_r, \mmusb \right)$ pair is outside the area defined by the data (points in each panel of Figure \ref{fig:depthcorr}), we assume the correction is equal to the value of the correction at the nearest boundary defined by the projection of the data in the measured-correction subset onto the best-fit plane (that is, we do not extrapolate). 

In addition to the release of classification data described in Section \ref{sec:release} above, we additionally present these ``corrected'', weighted classifications for each of the 8,130 subjects with deep exposures but for which we do not also have separate wide-field depth classifications, as well as the measured wide-field depth classifications for the measured-correction subset, for a total of 10,648 morphological classifications of deep-field subjects corrected to the wide-field average depth.

For these subjects, the wide-field depth classifications are given in Table \ref{table:data-depthcorr} and as a separate table in the data catalogs. The naming of columns is as described in Section \ref{sec:release}, except with an additional {\small \tt \_deepcorr} added to each relevant weighted-classification column. For example, the wide-field vote fraction for classifiers indicating an answer of `Smooth' to Task T00 is labelled {\small \tt t00\_smooth\_or\_featured\_a0\_smooth\_weighted\_frac\_deepcorr}, which is depth-corrected from the deep-exposure classification indicated in the {\small \tt t00\_smooth\_or\_featured\_a0\_smooth\_weighted\_frac} column. For those investigating science questions where it is advantageous to consider classifications from images of comparable depth across an entire sample \citep[e.g.,][]{bamford08}, we recommend using the {\small \tt \_deepcorr} classifications for subjects in the ``deep'' fields.

\begin{table*}
\resizebox{\textwidth}{!}{%
\begin{tabular}{@{}lccccccc}
\hline
\multicolumn{1}{l}{Data Column Name} &
\multicolumn{6}{c}{Subjects} &
\multicolumn{1}{c}{}
\\
\hline
\hline
ID & GDS\_12132 & GDS\_15834 & GDS\_17388 & GDS\_17613 & GDS\_20321 & GDS\_8970 & \\
RA & 53.099782 & 53.075878 & 53.126442 & 53.192374 & 53.159091 & 53.142340 & \\
Dec & -27.799512 & -27.768666 & -27.756544 & -27.752376 & -27.728037 & -27.827641 & \\
t00\_smooth\_or\_featured\_a0\_smooth\_weighted\_frac\_deepcorr & 0.67 & 0.55 & 0.37 & 0.70 & 0.36 & 0.64 & \\
t00\_smooth\_or\_featured\_a1\_features\_weighted\_frac\_deepcorr & 0.05 & 0.16 & 0.47 & 0.07 & 0.18 & 0.18 & \\
t00\_smooth\_or\_featured\_a2\_artifact\_weighted\_frac\_deepcorr & 0.28 & 0.28 & 0.16 & 0.23 & 0.45 & 0.18 & \\
t01\_how\_rounded\_a0\_completely\_weighted\_frac\_deepcorr & 0.84 & 0.32 & 0.87 & 0.59 & 0.26 & 0.43 & \\
t01\_how\_rounded\_a1\_inbetween\_weighted\_frac\_deepcorr & 0.15 & 0.67 & 0.13 & 0.36 & 0.72 & 0.51 & \\
t01\_how\_rounded\_a2\_cigarshaped\_weighted\_frac\_deepcorr & 0.01 & 0.01 & 0.00 & 0.05 & 0.02 & 0.06 & \\
t02\_clumpy\_appearance\_a0\_yes\_weighted\_frac\_deepcorr & 1.00 & 0.48 & 0.41 & 0.74 & 0.58 & 0.48 & \\
t02\_clumpy\_appearance\_a1\_no\_weighted\_frac\_deepcorr & 0.00 & 0.52 & 0.59 & 0.26 & 0.42 & 0.52 & \\
t03\_how\_many\_clumps\_a0\_1\_weighted\_frac\_deepcorr & 1.00 & 0.06 & 0.55 & 0.28 & 0.03 & 0.25 & \\
t03\_how\_many\_clumps\_a1\_2\_weighted\_frac\_deepcorr & 0.00 & 0.50 & 0.00 & 0.09 & 0.33 & 0.03 & \\
t03\_how\_many\_clumps\_a2\_3\_weighted\_frac\_deepcorr & 0.00 & 0.14 & 0.03 & 0.25 & 0.26 & 0.03 & \\
t03\_how\_many\_clumps\_a3\_4\_weighted\_frac\_deepcorr & 0.00 & 0.02 & 0.07 & 0.02 & 0.02 & 0.22 & \\
t03\_how\_many\_clumps\_a4\_5\_plus\_weighted\_frac\_deepcorr & 0.00 & 0.07 & 0.35 & 0.20 & 0.04 & 0.04 & \\
t03\_how\_many\_clumps\_a5\_cant\_tell\_weighted\_frac\_deepcorr & 0.00 & 0.21 & 0.00 & 0.16 & 0.32 & 0.43 & \\
t04\_clump\_configuration\_a0\_straight\_line\_weighted\_frac\_deepcorr & 0.00 & 0.02 & 0.00 & 0.02 & 0.00 & 0.00 & \\
t04\_clump\_configuration\_a1\_chain\_weighted\_frac\_deepcorr & 0.00 & 0.11 & 0.00 & 0.34 & 0.00 & 0.07 & \\
t04\_clump\_configuration\_a2\_cluster\_or\_irregular\_weighted\_frac\_deepcorr & 0.00 & 0.87 & 0.60 & 0.63 & 1.00 & 0.62 & \\
t04\_clump\_configuration\_a3\_spiral\_weighted\_frac\_deepcorr & 0.00 & 0.00 & 0.40 & 0.01 & 0.00 & 0.31 & \\
t05\_is\_one\_clump\_brightest\_a0\_yes\_weighted\_frac\_deepcorr & 0.00 & 0.25 & 0.40 & 0.42 & 0.49 & 0.38 & \\
t05\_is\_one\_clump\_brightest\_a1\_no\_weighted\_frac\_deepcorr & 0.00 & 0.75 & 0.60 & 0.58 & 0.51 & 0.62 & \\
t06\_brightest\_clump\_central\_a0\_yes\_weighted\_frac\_deepcorr & 0.00 & 0.28 & 1.00 & 1.00 & 0.51 & 0.18 & \\
t06\_brightest\_clump\_central\_a1\_no\_weighted\_frac\_deepcorr & 0.00 & 0.72 & 0.00 & 0.00 & 0.49 & 0.82 & \\
t07\_galaxy\_symmetrical\_a0\_yes\_weighted\_frac\_deepcorr & 0.00 & 0.04 & 0.66 & 0.05 & 0.00 & 0.34 & \\
t07\_galaxy\_symmetrical\_a1\_no\_weighted\_frac\_deepcorr & 1.00 & 0.96 & 0.34 & 0.95 & 1.00 & 0.66 & \\
t08\_clumps\_embedded\_larger\_object\_a0\_yes\_weighted\_frac\_deepcorr & 0.61 & 0.36 & 0.66 & 0.48 & 0.35 & 0.17 & \\
t08\_clumps\_embedded\_larger\_object\_a1\_no\_weighted\_frac\_deepcorr & 0.39 & 0.64 & 0.34 & 0.52 & 0.65 & 0.83 & \\
t09\_disk\_edge\_on\_a0\_yes\_weighted\_frac\_deepcorr & 0.00 & 0.06 & 0.05 & 0.00 & 0.72 & 0.40 & \\
t09\_disk\_edge\_on\_a1\_no\_weighted\_frac\_deepcorr & 0.00 & 0.94 & 0.95 & 0.00 & 0.28 & 0.60 & \\
t10\_edge\_on\_bulge\_a0\_yes\_weighted\_frac\_deepcorr & 0.00 & 0.00 & 1.00 & 0.00 & 1.00 & 0.50 & \\
t10\_edge\_on\_bulge\_a1\_no\_weighted\_frac\_deepcorr & 0.00 & 0.00 & 0.00 & 0.00 & 0.00 & 0.50 & \\
t11\_bar\_feature\_a0\_yes\_weighted\_frac\_deepcorr & 0.00 & 0.01 & 0.05 & 0.00 & 0.00 & 0.00 & \\
t11\_bar\_feature\_a1\_no\_weighted\_frac\_deepcorr & 0.00 & 0.99 & 0.95 & 0.00 & 1.00 & 1.00 & \\
t12\_spiral\_pattern\_a0\_yes\_weighted\_frac\_deepcorr & 0.00 & 0.00 & 0.29 & 0.00 & 1.00 & 0.00 & \\
t12\_spiral\_pattern\_a1\_no\_weighted\_frac\_deepcorr & 0.00 & 1.00 & 0.71 & 0.00 & 0.00 & 1.00 & \\
t13\_spiral\_arm\_winding\_a0\_tight\_weighted\_frac\_deepcorr & 0.00 & 0.00 & 0.83 & 0.00 & 1.00 & 0.00 & \\
t13\_spiral\_arm\_winding\_a1\_medium\_weighted\_frac\_deepcorr & 0.00 & 0.00 & 0.17 & 0.00 & 0.00 & 0.00 & \\
t13\_spiral\_arm\_winding\_a2\_loose\_weighted\_frac\_deepcorr & 0.00 & 0.00 & 0.00 & 0.00 & 0.00 & 0.00 & \\
t14\_spiral\_arm\_count\_a0\_1\_weighted\_frac\_deepcorr & 0.00 & 0.00 & 0.17 & 0.00 & 0.00 & 0.00 & \\
t14\_spiral\_arm\_count\_a1\_2\_weighted\_frac\_deepcorr & 0.00 & 0.00 & 0.68 & 0.00 & 0.00 & 0.00 & \\
t14\_spiral\_arm\_count\_a2\_3\_weighted\_frac\_deepcorr & 0.00 & 0.00 & 0.00 & 0.00 & 0.00 & 0.00 & \\
t14\_spiral\_arm\_count\_a3\_4\_weighted\_frac\_deepcorr & 0.00 & 0.00 & 0.17 & 0.00 & 0.00 & 0.00 & \\
t14\_spiral\_arm\_count\_a4\_5\_plus\_weighted\_frac\_deepcorr & 0.00 & 0.00 & 0.00 & 0.00 & 0.00 & 0.00 & \\
t14\_spiral\_arm\_count\_a5\_cant\_tell\_weighted\_frac\_deepcorr & 0.00 & 0.00 & 0.00 & 0.00 & 1.00 & 0.00 & \\
t15\_bulge\_prominence\_a0\_no\_bulge\_weighted\_frac\_deepcorr & 0.00 & 0.75 & 0.00 & 0.00 & 0.00 & 1.00 & \\
t15\_bulge\_prominence\_a1\_obvious\_weighted\_frac\_deepcorr & 0.00 & 0.20 & 0.62 & 0.00 & 1.00 & 0.00 & \\
t15\_bulge\_prominence\_a2\_dominant\_weighted\_frac\_deepcorr & 0.00 & 0.05 & 0.38 & 0.00 & 0.00 & 0.00 & \\
t16\_merging\_tidal\_debris\_a0\_merging\_weighted\_frac\_deepcorr & 0.01 & 0.10 & 0.35 & 0.09 & 0.15 & 0.54 & \\
t16\_merging\_tidal\_debris\_a1\_tidal\_debris\_weighted\_frac\_deepcorr & 0.02 & 0.02 & 0.01 & 0.07 & 0.10 & 0.04 & \\
t16\_merging\_tidal\_debris\_a2\_both\_weighted\_frac\_deepcorr & 0.01 & 0.01 & 0.04 & 0.02 & 0.05 & 0.03 & \\
t16\_merging\_tidal\_debris\_a3\_neither\_weighted\_frac\_deepcorr & 0.96 & 0.87 & 0.60 & 0.82 & 0.70 & 0.39 & \\
\hline
\end{tabular}}
\caption{Depth-corrected classifications for the ``measured-correction'' sample defined in Section \ref{sec:depth}. The complete version of this table is available in electronic form and at http://data.galaxyzoo.org. The printed table shows a transposed subset of the full table to illustrate its format and content.}
\label{table:data-depthcorr}
\end{table*}

\subsection{Resolution effects}\label{sec:resolution}

In addition to variations in classification as a function of image depth discussed above, the minimum resolved physical scale in a galaxy changes as a function of redshift, which affects the detectability of smaller-scale features. However, at the redshifts where the bulk of galaxies in the Galaxy Zoo CANDELS data set lie ($z \gtrsim 0.5$), the redshift dependence of the angular diameter distance is relatively flat compared to its evolution at lower redshifts. This means that the physical resolution changes only slightly over the bulk of the survey. At $z > 1$, where Galaxy Zoo CANDELS adds substantially new rest-frame optical morphologies compared to previous \emph{HST} morphological surveys \citep[e.g.,][Willett et al., in preparation]{scarlata07}, the maximum variation in physical resolution as a function of redshift is approximately 5 per cent. Given the resolution of the drizzled \emph{HST} images (Section \ref{sec:images}) and a PSF FWHM $\approx 2.5$ pixels, a morphological feature in a galaxy must be larger than approximately 1 kpc to be resolved in $F160W$ at any redshift covered by Galaxy Zoo CANDELS. 

More specifically, the FWHM of the WFC3 PSF is equivalent to 0.92 kpc at $z = 0.5$; this increases to 1.22 kpc at $z = 1$ and  1.27 kpc at $z = 1.6$, the redshift at which physical resolution is at its worst. At the redshift where the central wavelength of the $F160W$ filter is approximately aligned with the rest-frame $B$ band, $z=2.7$, the FWHM of the WFC3 PSF in physical units is 1.19 kpc. The images used here therefore cannot resolve intrinsically distinct features smaller than $\sim 1-1.3$~kpc, so a galaxy with \emph{only} features smaller than this is likely to be classified as ``smooth'' in Galaxy Zoo CANDELS. There may be exceptions due to the fact that the ACS PSF is approximately half the size, and the colour images use ACS $F814W$ images in the blue channel. A galaxy with features of a size just below the detection limit in WFC3 \emph{but} which are also bright in $F814W$ (which is in the rest-frame UV for $z \gtrsim 1$) may show blue resolved features in the subject images.

ACS and WFC3 on \emph{HST} provide the highest-resolution images currently available. These images may in the future be used in comparison with morphological studies using the \emph{James Webb Space Telescope}, although that telescope is not as well optimised for surveys as \emph{HST}. The classifications presented here will likely be of substantial use to large-scale morphological studies using the \emph{Euclid} mission \citep{refregier10}, which will cover more of the sky at approximately half the resolution of \emph{HST} at any wavelength. For now, however, we note going forward that the features reported by Galaxy Zoo CANDELS will in general be limited to those with a physical size of at least $\sim 1$ kpc.

%%%%%%%%%%%%%%%%%%%%%%%%%%%%%%%%%%%%%%%%%%%%%%
%
%  
\section{Comparison to other visual classifications}\label{sec:comparison}
%
%
%%%%%%%%%%%%%%%%%%%%%%%%%%%%%%%%%%%%%%%%%%%%%%

Most of the galaxies in the CANDELS data set have additional visual classifications available in the form of expert classifications from astronomers and undergraduate students who are members of the CANDELS team. Analysis of the full set of classifications in that separate project is still underway; the first release of classifications from the GOODS-South field is presented by \citet{kartaltepe15}, hereafter K15, who also detail the project design and objectives, including the classification interface. Consensus classifications from the UDS field are also available (Kartaltepe et al., in preparation). For each galaxy in all fields, between 3 and 7 (typically 3) members of the CANDELS team provided classifications.

The classification scheme described in K15 is substantially different to that presented here. Firstly, while that project collects detailed classifications about a number of possible structural features (with 37 different responses possible), they do not always align precisely with the questions asked in Galaxy Zoo CANDELS. For example, the Main Morphology Class of K15 requires the classifier to select at least one option from among ``disk'', ``spheroid'', and ``peculiar/irregular'' galaxy types, along with options for ``point source/compact'' and ``unclassifiable''. The last of these is not an option Galaxy Zoo provides, and the first two are not necessarily the same as task T00's responses of ``features'' versus ``smooth'' \citep[note: this also means we cannot compare to the machine classifications of][as there are no categories in that study that translate to the measurements made by Galaxy Zoo volunteers]{huertascompany15}. While Galaxy Zoo does classify bulge strength, it does so after multiple branches of the decision tree, and therefore this is not easily comparable to a first-tier task.

In fact, \emph{all} responses collected by the CANDELS team interface are first-tier tasks: the classifier is presented with all 37 options at once. Additionally, colour composites are not used in that project. Images from each ACS and WFC3 filter are presented separately within the interface, with options for the classifier to specify when classifications differ significantly between filters. Classifiers may also view the segmentation map in the $F160W$ band, and in the Perl/DS9 version of the CANDELS team interface the classifier may adjust the stretch of the image. These options are not available to Galaxy Zoo classifiers. On the other hand, the Galaxy Zoo decision tree asks multiple questions designed to elucidate the spatial configuration of clumps in a galaxy, whereas the CANDELS team interface instead requests a clumpiness rating and makes a distinction between patchiness and clumpiness. 

The CANDELS team visual classifications in the GOODS-South and UDS fields have been used in multiple studies, many of which adapt new metrics based on the raw consensus classifications \citep{guo15, rosario15}. For example, \citet{guo15} combine the raw classifications of clumpiness and patchiness to create single-parameter measures of each, while McIntosh et al. (in preparation) apply a user weighting scheme to the K15 classifications and produce a ``Diskiness'' metric $D_v$, ranging from 0 (no disk) to 1 (pure disk). In the Galaxy Zoo CANDELS classifications each question about clumpiness is kept distinct, and there is no single question that identifies a disk independent of other features.

Despite these significant differences, it is nevertheless helpful to compare the CANDELS team classifications to the Galaxy Zoo CANDELS classifications. There are 15,383 Galaxy Zoo CANDELS subjects which also have classifications from the CANDELS team. Figure \ref{fig:candels_compare} shows the comparison of consensus vote fractions in four categories: Featured, Merger or Interaction, Edge-On, and Spiral. Figure \ref{fig:candels_compare_diskiness} compares the CANDELS team diskiness parameter $D_v$  (McIntosh et al, in preparation) to the Galaxy Zoo CANDELS ``Features or Disk'' answer to task T00. For all comparisons below we have compared the subset of CANDELS sources that: are brighter than the surface brightness limit $\mmusb < 24.5$, have visual classifications from both teams, have \emph{not} been deemed ``unclassifiable'' by the CANDELS team ($f_{\rm unc,~\textsc{ct}} < 0.3$), and have \emph{not} been rejected as stars or artifacts by more than 50\% of classifiers for either project. This selection results in a sample of 13,145 galaxies. The surface brightness limit is chosen to favour inclusiveness. While this choice adds somewhat to the noise seen in Figures \ref{fig:candels_compare} and \ref{fig:candels_compare_diskiness}, we do not expect it to bias the correlations, as each classification project uses visual classifications of the same data so should be equally affected (or unaffected) by surface brightness issues.
%described in Section \ref{sec:sb} 
  
%%%%% [FIGURE: CANDELS and GZ Comparisons] %%%%%
\begin{figure*}
\includegraphics[scale=0.88]{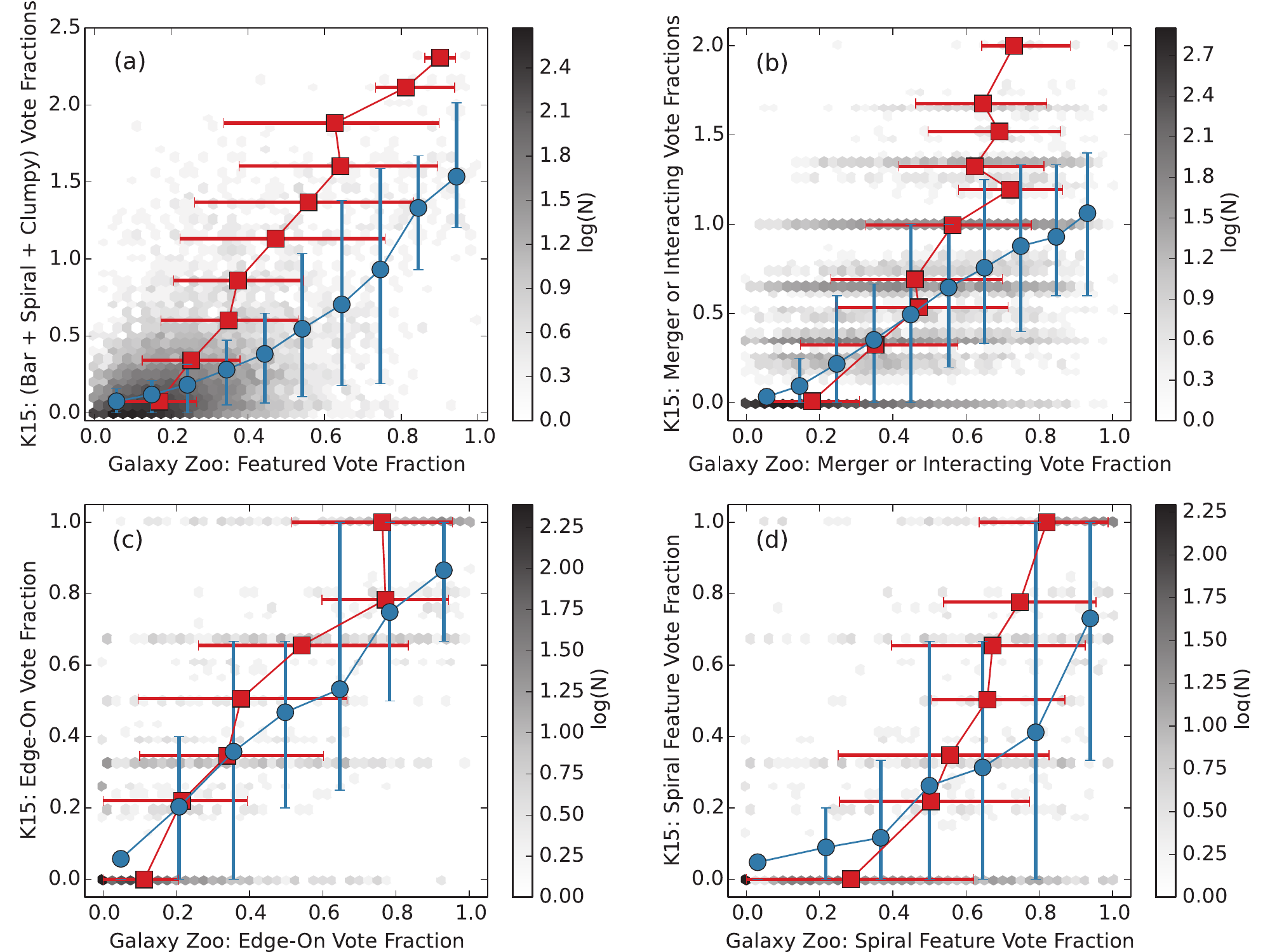}
\caption{
Comparison of Galaxy Zoo classifications with visual classifications from the CANDELS team \citep[][K15]{kartaltepe15}. The classification questions differ between the two projects, but we have selected 4 different classifications which are the most similar: (a) the sum of vote fractions in K15 for spiral, bar, and clumpy features, versus the Galaxy Zoo vote fraction for ``features or disk'' in task T00. Note that the sum of these vote fractions from K15 can add to $> 1$; (b) vote fractions for merger or interactions (task T16 in the Galaxy Zoo decision tree) for those subjects not identified as ``star or artifact'' in task T00; (c) vote fractions for the presence of an edge-on disk (task T09) for subjects that are neither artifacts nor predominantly smooth, nor dominated by clumps; and (d) vote fractions for the presence of spiral arms (task T12) for those subjects in panel (c) that are not edge-on. In all panels, the number of individual galaxies in a given hexagon in parameter space is shown by its shaded value. Red squares show the average Galaxy Zoo vote fraction binned by CANDELS team classification; Blue circles show the average CANDELS team classification in bins of Galaxy Zoo vote fraction. Error bars on red and blue points show the region enclosing the middle 68 per cent of values in that bin. When parameters are chosen that measure similar features between the different visual classification methods, they track each other well across many different kinds of structural classification.
}
\label{fig:candels_compare}
\end{figure*}
%%%%% END FIGURE %%%%%

%%%%% [FIGURE: CANDELS and GZ Comparisons] %%%%%
\begin{figure}
\includegraphics[scale=0.7]{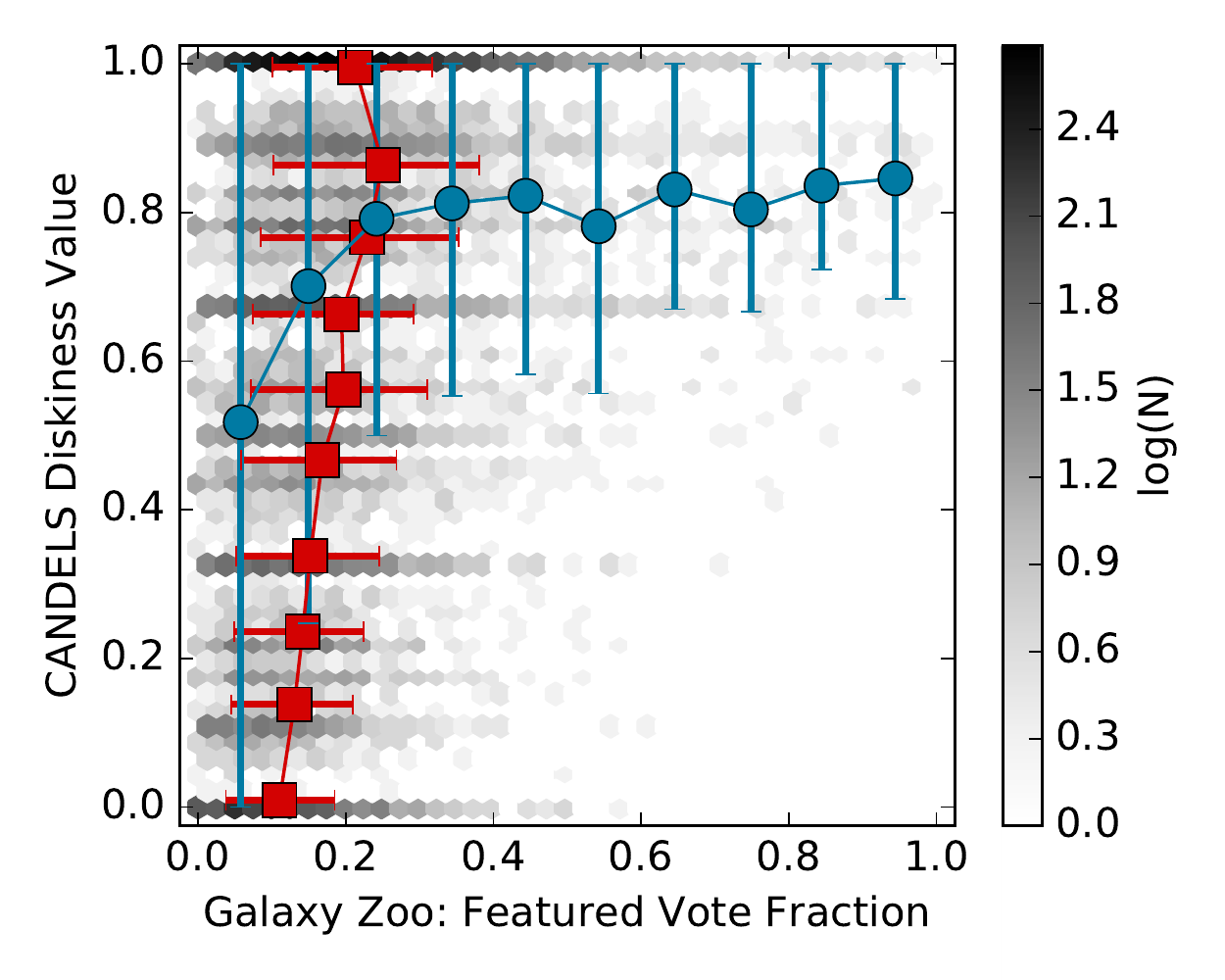}
\caption{
Comparison of Galaxy Zoo ``featured'' classifications with visually-determined Diskiness parameter $D_v$ from the CANDELS team (McIntosh et al, in preparation). The number of individual galaxies in a given hexagon in parameter space is shown by its shaded value. Red squares show the average Galaxy Zoo vote fraction binned by CANDELS team classification; Blue circles show the average CANDELS team classification in bins of Galaxy Zoo vote fraction. Error bars on red and blue points show the region enclosing the middle 68 per cent of values in that bin. While these two parameters weakly correlate ($\rho = 0.29$), they do not in general measure the same thing. Galaxies with a high \ffeat\ value are generally disky ($\left< D_v \right> \sim 0.8$ for galaxies with $\mffeat > 0.5$, though within this subset the parameters are uncorrelated), but Galaxy Zoo volunteers also identify mergers and other features not associated with disks. As these are rare, one can select a sample of disk galaxies based on a \ffeat\ threshold alone, but such a selection will also remove a potentially important sample of relatively featureless disks.
}
\label{fig:candels_compare_diskiness}
\end{figure}
%%%%% END FIGURE %%%%%

\subsection{Featured Galaxies}\label{sec:compare_featured}

We seek to compare the overall classification of CANDELS galaxies as ``smooth'' or ``featured'' between Galaxy Zoo and the CANDELS team. However, the team interface described in K15 does not specifically ask about this distinction. It does ask about disks and spheroids; however, equating ``smooth'' to ``spheroid'' and ``featured'' to ``disk'' requires assumptions about galaxies at $z > 1$ that we would prefer to avoid (see Sections \ref{sec:feat_not_disk} and \ref{sec:result} for further discussion of ``smooth'' disks).

In order to compare the ``featured'' vote fraction for T00 in Galaxy Zoo CANDELS to a more similar measure from the CANDELS team visual classifications, we construct a CANDELS-team ``featured'' galaxy measure using a combination of vote fractions from the CANDELS team classifications (K15). We choose a set of structures that are unambiguously inconsistent with a smooth light distribution in a galaxy, namely spiral arms, clumps, and bar features. 

Within the CANDELS team classification interface, a classifier may indicate the presence of a bar or spiral arms by selecting one response for each, $f_{\rm bar, \textsc{ct}}$ or $f_{\rm spiral, \textsc{ct}}$. The clumpy classification, however, is actually a rating of both ``clumpiness'' and ``patchiness'', in a $3 \times 3$ grid with ratings from 0 to 2 along each axis. According to K15, ``Clumps are concentrated independent knots of light while patches are more diffuse structures.'' Both are inconsistent with a smooth light distribution, so we include both in the creation of a ``featured'' vote for the CANDELS team classifications.

We combine clumpy classifications within this matrix of possible responses into a single value, following an approach similar to \citet{guo15} but modified to include clumpiness and patchiness in the same metric. Each vote is weighted by the strength of features that it indicates, by assigning a weight of 0.25 for each level along each axis. For a clumpiness rating $i$ and a patchiness rating $j$, the weight for that vote fraction is $$w_{ij} = 0.25 (i + j) .$$ For example, the weight for $C_1 P_2$ = 0.75. As the maximum value within the selection matrix is  $C_2 P_2$, the maximum weight is 1. The overall clumpy vote fraction for a given object is then $$f_{\rm clumpy, \textsc{ct}} = \sum_i{\sum_j{w_{ij} f_{ij}}} .$$ We note that classifiers may make multiple selections within the clumpiness/patchiness rating matrix, so the weighted, summed vote fraction $f_{\rm clumpy}$ can in principle exceed 1.

Figure \ref{fig:candels_compare}a shows the summed ``featured'' vote fraction for the CANDELS team, $f_{\rm bar, \textsc{ct}} + f_{\rm spiral, \textsc{ct}} + f_{\rm clumpy, \textsc{ct}}$, versus the Galaxy Zoo vote fraction for the response ``features or disk'' to task T00, for the 13,145 galaxies that have been classified by both and that meet the surface brightness and other criteria described at the start of Section \ref{sec:comparison}. Figure~\ref{fig:candels_compare} shows the 2-D histogram via hexagonal shading, indicating that in both projects a high vote for features of any kind is relatively rare (most galaxies have $\mffeat \sim 0$ and $f_{\rm features, \textsc{ct}} \sim 0$). To guide the eye, Figure \ref{fig:candels_compare} also shows binned averages: blue circles show the average CANDELS team classification in equal-sized bins of Galaxy Zoo vote fraction, while red squares show the average Galaxy Zoo classification binned by CANDELS team classification. 

The featured vote fractions track each other well, with a clear and highly significant positive correlation (Spearman $\rho = 0.45$, $p < 2 \times 10^{-16}$\footnote[\dagger]{The reported $p$ value is consistent with zero within machine precision, i.e., highly significant.}). There are virtually no galaxies for which the CANDELS team voted strongly for features being present but the Galaxy Zoo classifiers did not. It is also rare for the Galaxy Zoo classifiers to find a proportionally higher vote fraction for features than the CANDELS team, although the few examples seen in this parameter space may contain examples of distraction bias in a classification interface that presents dozens of choices simultaneously \citep[e.g.,][]{simons99,iyengar00}. This is clearly a small effect, however: on the whole the classifications agree remarkably well with each other.

\subsection{Merging or Interacting Galaxies}

The final classification task (T16) in the Galaxy Zoo decision tree asks whether the classifier sees evidence of a merger, or of tidal interaction, or both, or neither. The CANDELS team Interaction Class (K15) asks the classifier to decide between the following classes: merger, interaction (within segmap), interaction (outside segmap), non-interacting companion, and none. There is also a separate flag within the CANDELS team classification to indicate whether a galaxy has tidal arms. Because these selections between projects are similar but not exactly the same, we choose to compare the sum of all signs of interaction of any kind within both projects. Specifically, we consider the sum of vote fractions within the CANDELS team classifications for ``merger'', ``interaction within segmap'',  ``interaction beyond segmap'', and ``tidal arms'', while for Galaxy Zoo we consider the sum of vote fractions for ``merging'', ``tidal debris'', or ``both''. For Galaxy Zoo the maximum vote fraction is 1, whereas the maximum value for the combined CANDELS team vote is 2.

Figure \ref{fig:candels_compare}b compares these fractions for each galaxy in the same way and using the same sample as Figure \ref{fig:candels_compare}a, with darker shaded bins representing a higher number of galaxies within that bin, and with red squares and blue circles indicating averages of one classification binned by the other, as described in Section \ref{sec:compare_featured} above. The striations seen in the hexagonal bins reflect the finite number of possible vote fractions within the CANDELS team votes; this structure was not seen in Figure \ref{fig:candels_compare}a due to the weighted combination of clumpy vote fractions. 

Although Galaxy Zoo and the CANDELS team measure different aspects of mergers differently, in combination the merger/interaction vote fractions clearly correlate ($\rho = 0.67$, $p < 2 \times 10^{-16}$). The correlation is likewise strong when we compare Galaxy Zoo vote fractions to the combined and re-normalised merger/interaction value from \citet{rosario15} ($\rho = 0.70$, $p < 2 \times 10^{-16}$; note that we show the raw vote fraction combination in Figure \ref{fig:candels_compare}b for consistency with the other figure panels). As in the comparison between overall featured fractions, there are more examples where the Galaxy Zoo vote fraction is notably higher than the CANDELS team vote fraction than vice-versa. Examination of galaxies where Galaxy Zoo $f_{\rm merger\ or\ interaction} > 0.5$ and CANDELS team $f_{\rm merger\ or\ interaction, \textsc{ct}} = 0$ indicates some cases where a merger or tidal feature is clearly present, but others where it is less obvious whether a nearby companion is interacting. 

Indeed, among this sub-sample the CANDELS team vote fraction for ``non-interacting companion'' is considerably higher on average than for the overall sample. This option is not explicitly available to Galaxy Zoo classifiers, although even moderately experienced classifiers, particularly those who participate in discussions within the community Talk software, will in general select ``neither'' if they decide the companion is not interacting. This explains why the number of galaxies showing this mismatch is much smaller (less than 2\% of the sample) than the overall number of galaxies which CANDELS team classifications mark as having a non-interacting companion. Future analyses of mergers and interacting galaxies may find a combination of Galaxy Zoo and CANDELS team classifications useful for eliminating the effects of distraction bias and distinguishing between interacting and non-interacting companions.

\subsection{Edge-On Galaxies}

As described in Section \ref{sec:usage}, the branched nature of the Galaxy Zoo decision tree means that selecting a sample for comparison of edge-on vote fraction requires care. We thus consider, in addition to the previous sample requirements, that a galaxy must also have a featured vote fraction $\mffeat \geq 0.3$ and a not-clumpy vote fraction $f_{\rm not\ clumpy} \geq 0.3$. This selection favours completeness over purity, and is thus appropriate for a comparison of different visual classification methods. The selection results in a sample of 1,611 galaxies.

Both the CANDELS team and Galaxy Zoo classifications allow for the flagging of a galaxy as edge-on with a single selection, facilitating a direct comparison. Figure \ref{fig:candels_compare}c shows the CANDELS team versus the Galaxy Zoo classification vote fractions. The two agree very well ($\rho = 0.72$, $p < 2 \times 10^{-16}$), with the average vote fractions in one classification schema binned by the other (red squares and blue circles) generally consistent with a 1:1 line.

\subsection{Spiral Galaxies}

The spiral galaxy tasks in Galaxy Zoo are one branch below the edge-on disk galaxy task (T09), introducing another dependency on the sample selection, as described in Section \ref{sec:usage}. From within the sample used to construct Figure \ref{fig:candels_compare}c, we further require a vote of $f_{\rm not\ edge-on} \geq 0.5$, a selection chosen to balance the desire for completeness with the need to be able to see spiral arms if they are present. This selects 1,192 galaxies, whose positions in Figure \ref{fig:candels_compare}d are shown in the 2-D shaded histograms. 

As in all other morphological parameters shown in Figure \ref{fig:candels_compare}, the visual classifications from both projects agree very well ($\rho = 0.61$, $p < 2 \times 10^{-16}$). Outliers in this figure include some examples that are best explained by distraction bias in the absence of a decision tree in CANDELS, and also include a few examples of distraction bias of a different sort in Galaxy Zoo: a handful of subjects that include a spiral galaxy very near the central, much fainter, galaxy. Such examples are a very small part of the overall sample, and are relatively easily rejected from a sample selection in any case. We also note that those galaxies with $f_{\rm spiral} > 0.5$ have a CANDELS ``diskiness'' parameter mean value of $D_v \sim 0.8$: that is, where a (weighted) majority of Galaxy Zoo classifiers indicated the presence of spiral arms, a high fraction of CANDELS classifiers identified clear visual signs of a disk. 

In general the classifications agree very well along this and other morphological axes which are directly comparable between the CANDELS team and Galaxy Zoo visual classifications. In most cases, however, the Galaxy Zoo and CANDELS team visual classifications are complementary rather than directly translatable. This is potentially an advantage for those wishing to include both in more complex selections. For instance, a combination of Galaxy Zoo and CANDELS team visual classifications would likely be helpful in selecting a sample of interacting galaxies with the optimum combination of completeness and purity, while a combination of clumpy selections from each would allow for a comparison of clumpiness versus patchiness in galaxies with a range of clump configurations.

\subsection{``Disk-like'' and ``Featured'' are not equivalent}\label{sec:feat_not_disk}

Using a different combination of CANDELS team classifications and Galaxy Zoo classifications, we can directly test how well galaxy features correlate with visually identified disks by comparing the ``diskiness'' parameter $D_v$ (McIntosh et al, in preparation) assessed by CANDELS team members to the Galaxy Zoo ``Features or disk'' vote fraction in Figure \ref{fig:candels_compare_diskiness}. The $D_v$ parameter is a visual assessment of light concentration intended to distinguish disks from spheroids even in the absence of features traditionally associated with disks, given high enough data quality. We follow McIntosh et al. in selecting only galaxies where there was high agreement of disk and/or spheroid nature ($DS_w > 0.65$ from that work), and galaxies with high scores for their quality metric ($Q_w > 0.65$) and low scores for their ``Unclassifiable'' measure ($U_w < 0.35$). This selection, in combination with the surface brightness and non-artifact criteria described above, results in a selection of 11,780 galaxies. Figure \ref{fig:candels_compare_diskiness} uses the same hexagonal binning for the 2-D histograms and binned averages as Figure \ref{fig:candels_compare}. 

Figure \ref{fig:candels_compare_diskiness} confirms that the Galaxy Zoo ``features or disk'' classification does not in general measure the same property as $D_v$. The two parameters are only weakly correlated (Spearman $\rho = 0.29, p < 2 \times 10^{-16}$), and at $\mffeat > 0.5$ there is no significant correlation, although $94$\% of galaxies with $\mffeat > 0.5$ have $D_v > 0.5$. In other words, those galaxies with very high featured vote fractions are also identified as disk-dominated galaxies by the CANDELS team. Indeed, visual inspection of these sources shows a high fraction of grand design spirals and other striking disk morphologies; the few subjects with high \ffeat\ and low $D_v$ are clearly highly complex systems, typically obvious interactions with tidal features or other strong asymmetries.

While it may thus be the case that the purity of a sample of disks selected via a threshold value of \ffeat\ varies directly with the threshold value chosen, Figure \ref{fig:candels_compare_diskiness} indicates that such a sample will not be complete for any reasonable \ffeat\ threshold. There is a high concentration of galaxies with $D_v \sim 1$ and $\mffeat < 0.3$, that is, galaxies which CANDELS team members identified visually as having a disk-like light profile but which Galaxy Zoo volunteers indicated are either smooth or have only weak features. We return to these potentially interesting galaxies in Section \ref{sec:result}, but for now note that task T00 in the Galaxy Zoo classification tree is intended to provide only a descriptive classification of whether a galaxy has features or not. As such, although \ffeat\ is useful in selecting disk galaxy samples, these samples may be biased against selection of featureless disks. Depending on the particular research question, a more complex selection or a correction for this effect \citep[e.g., that performed in][]{simmons14} may be necessary.

%%%%%%%%%%%%%%%%%%%%%%%%%%%%%%%%%%%%%%%%%%%%%%
%
%  
\section{The evolution of ``Smooth'' and ``Featured'' galaxies}\label{sec:result}
%
%
%%%%%%%%%%%%%%%%%%%%%%%%%%%%%%%%%%%%%%%%%%%%%%

%%%%% [FIGURE: dM] %%%%%
\begin{figure*}
\includegraphics[scale=0.85]{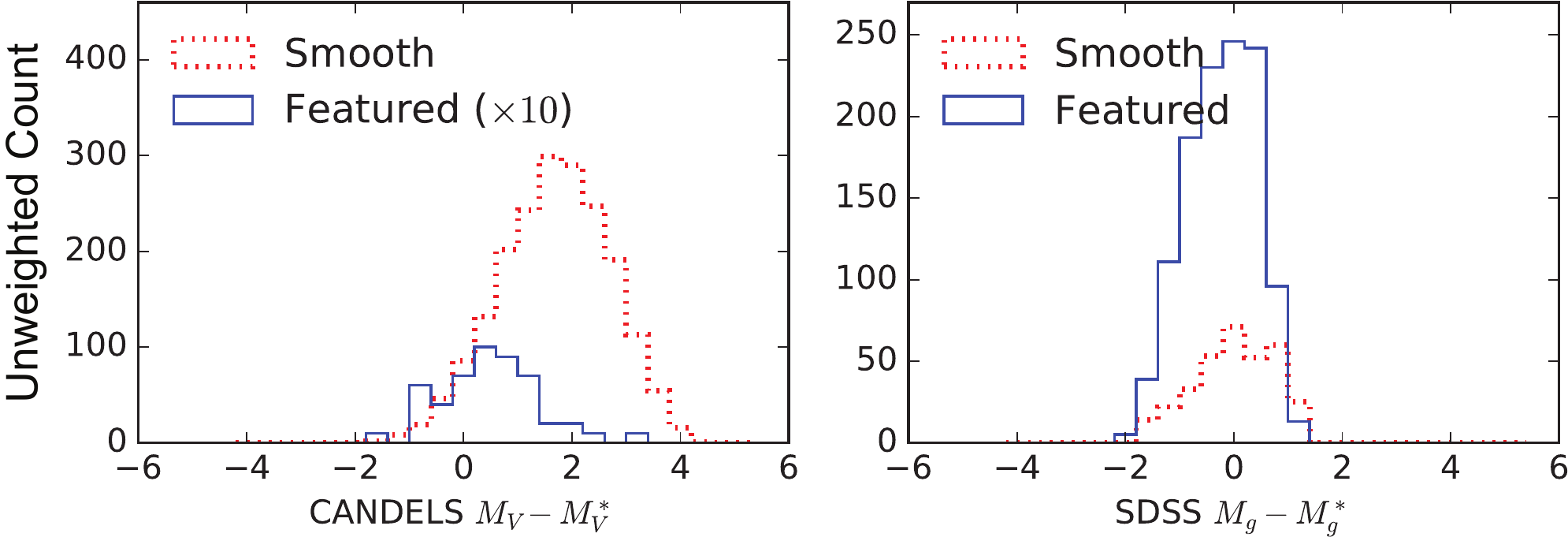}
\caption{
Distributions of luminosity function sampling for galaxy sub-samples selected by morphology and redshift. $M - M^*$ indicates where each galaxy lies on the rest-frame optical luminosity function for galaxies relative to $M^*$ at that redshift; values $>0$ indicate galaxies fainter than $M^*$. The left panel shows smooth (blue solid line) and featured (red dotted line) galaxies from Galaxy Zoo CANDELS at $1 \leq z \leq 3$ using rest-frame $V$-band absolute magnitudes and the luminosity function measured by \citet{marchesini12}. The featured sample is considerably smaller and brighter than the smooth sample (we show the featured distribution counts multiplied by $10$ for easier comparison). The right panel shows smooth and featured galaxies from Galaxy Zoo 2 \citep{willett13} using the rest-frame $g$-band absolute magnitudes and the luminosity function measured by \citet{blanton01}. Each sample has substantially different luminosities and luminosity distributions; further comparison of the samples requires controlling for these differences.}
\label{fig:dM}
\end{figure*}
%%%%% END FIGURE %%%%%

%%%%%%%%% [FIGURE: dM2] %%%%%%%%%
\begin{figure*}
\includegraphics[scale=0.85]{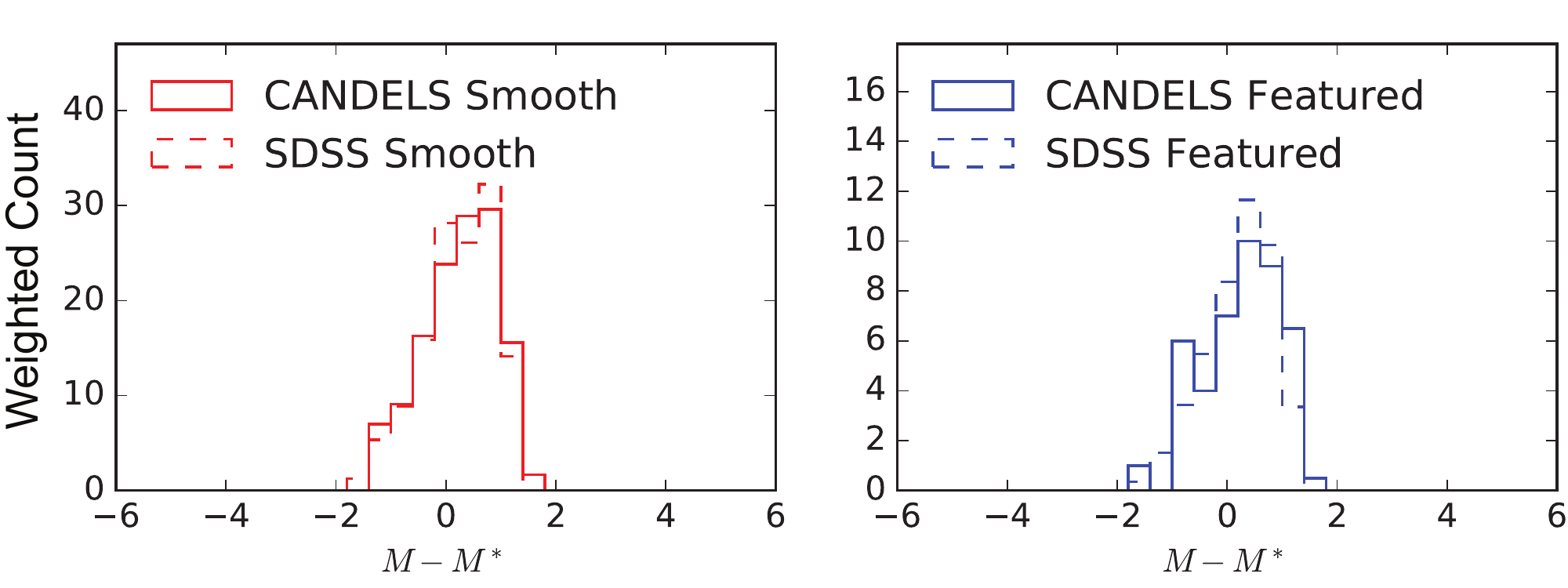}
\caption{
Weighted distributions of $M - M^*$ for the smooth (left panel) and featured (right panel) samples shown in figure \ref{fig:dM}. $M$ is the galaxy absolute rest-frame magnitude in $V$ (CANDELS) or $g$ (SDSS), and $M^*$ is the ``knee'' of the luminosity function in the relevant band, evaluated at the redshift of each observed galaxy. The weights for each galaxy in each bin are chosen so that all samples have statistically indistinguishable $M-M^*$ distributions. Weights are normalised separately for smooth and featured distributions and chosen so that the maximum weight for any single galaxy is 1.0; this results in effective counts of 132 galaxies per smooth sample and 44 galaxies per featured sample (CANDELS and SDSS are shown as solid and dashed curves, respectively). The peak of the distribution is slightly fainter than $M^*$ at each redshift, and the full weighted distribution includes galaxies at luminosities $\pm 2$ magnitudes from $M^*$. The bulge properties of these weighted samples are compared in Figure \ref{fig:btot}.
}
\label{fig:dM_wt}
\end{figure*}

%%%%% [FIGURE: BTot] %%%%%
\begin{figure*}
\includegraphics[scale=0.85]{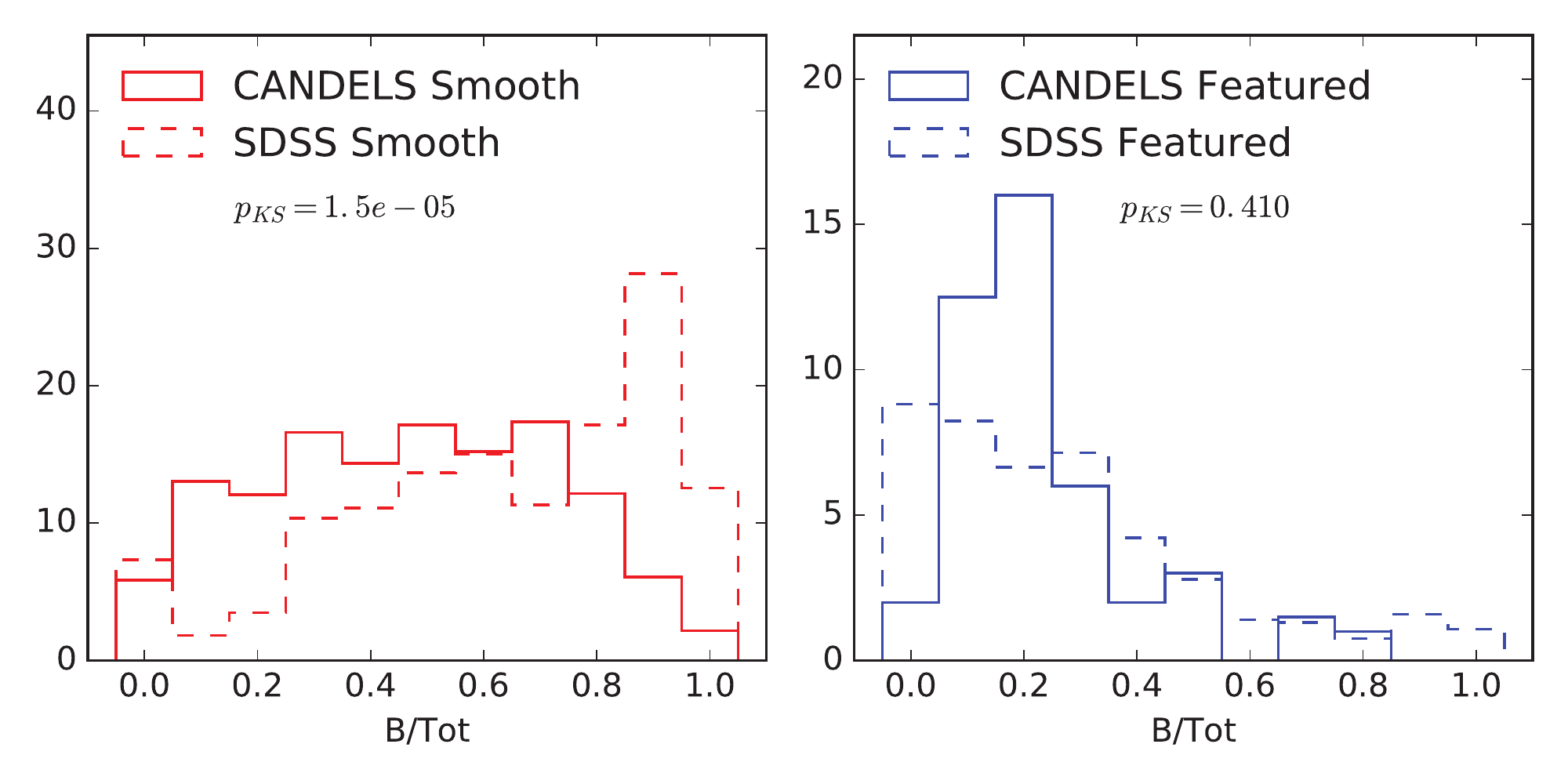}
\caption{
Distributions of bulge-to-total ratios for samples of galaxies observed by CANDELS (solid curves) and SDSS (dashed curves) that are unambiguously Smooth (red curves) and Featured (blue curves). The samples are weighted to control for differences in sampling of the luminosity functions at each redshift. The CANDELS galaxies span $1 \leq z \leq 3$; the SDSS galaxies are chosen to match the same range of physical resolutions in kpc given median seeing of $1^{\prime \prime}.4$ ($0.0418 \leq z \leq 0.0436$). The featured samples are generally disk-dominated, which is expected given that the majority of features measured by Galaxy Zoo 2 and Galaxy Zoo CANDELS are associated with disk instabilities. A K-S test between featured samples shows no statistically significant result. However, the Smooth galaxy samples have very different distributions between $1 \leq z \leq 3$ and $z \approx 0.04$. In particular, while the more local sample of smooth galaxies shows a high fraction of bulge-dominated galaxies, the distribution of higher-redshift ``smooth'' galaxies is relatively uniform and includes a substantial population of disk-dominated galaxies at $1 \leq z \leq 3$ which have no evidence of ``features'' often associated with disks. A K-S test between smooth samples returns $p < 1.5 \times 10^{-5}$ ($4.3 \sigma$). Smooth galaxies at $z \sim 0$ and $z \geq 1$ are most likely not drawn from the same underlying distribution of morphologies.
}
\label{fig:btot}
\end{figure*}
%%%%% END FIGURE %%%%%

%%%%% [FIGURE: BTot and colour for smooth CANDELS galaxies] %%%%%
\begin{figure*}
\includegraphics[scale=0.85]{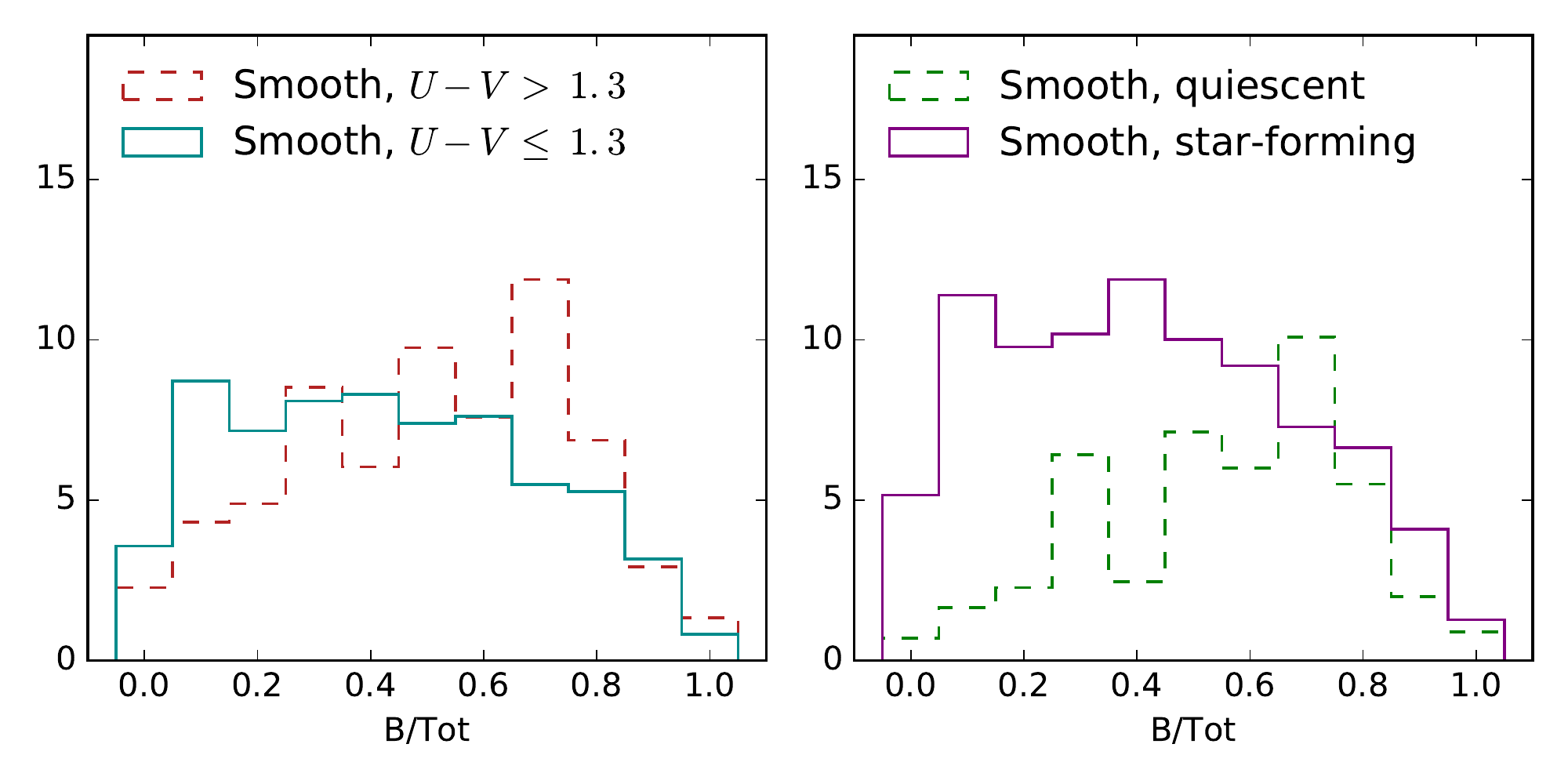}
\caption{
Distributions of bulge-to-total ratios for smooth galaxies in CANDELS with $1 \leq z \leq 3$. Distributions use the same weights as in Figures \ref{fig:dM_wt} and \ref{fig:btot}. In the left panel, smooth galaxies are sorted by rest-frame $U-V$ colour: those with $U-V > 1.3$ are shown in the red-dashed histogram, and those with $U-V \leq 1.3$ are shown in the solid cyan histogram. In the right panel, the smooth sample is split into quiescent (green-dashed) and star-forming (purple solid) sub-samples; these categories are applied via the criteria of \citet{williams09} (also shown as dashed lines in our Figure \ref{fig:UVJ}), which additionally incorporate $V-J$ colour. Neither $U-V$ nor $V-J$ colour alone are a useful means of separating disk-dominated from spheroid-dominated smooth galaxies in this sample. Two-colour information is more effective, but only partially so: using the level of star forming activity as a proxy for whether a smooth galaxy is disk-dominated or spheroid dominated can at most reach a completeness level of 77 per cent and a purity level of 63 per cent in this sample.
}
\label{fig:btot_colour}
\end{figure*}
%%%%% END FIGURE %%%%%

Although the morphological classifications described here present quantified visual assessments of a range of galaxy properties, the classification tree described in Section \ref{sec:tree} never explicitly asks the classifier to decide whether a galaxy has a disk. While many questions ask about disk instability features (or their absence), there is no single question that attempts to positively identify all disks. This choice of what \emph{not} to ask, which echoes that of previous Galaxy Zoo projects, partly reflects a discomfort with asking classifiers to assess light concentrations by eye without any further context. 

Thus within the Galaxy Zoo CANDELS classifications disks may be identified by the presence of specific features, but the absence of these features does not necessarily imply the lack of a disk, particularly at the epochs probed here. Decoupling questions about ``features'' from measures of a galaxy's diskiness enables these to be assessed independently. Specifically, while galaxies with strong ``features'' (as defined by the classification tree in Section \ref{sec:tree}) may be prone to systematic biases in identification of disks via \citet{sersic68} index measurements, galaxies which are more ``smooth'' do not suffer from these effects, and thus disk strength may be more accurately assessed. With the advent of new tools to measure light profiles via simultaneous consideration of multi-wavelength imaging \citep{haeussler13}, measurements of relative bulge and disk strengths are now possible with much higher accuracy than available for single-wavelength measurements at $z \sim 2$. 

To compare the Galaxy Zoo classifications of ``featured'' and ``smooth'' to measurements of disk strength, we first select samples of Smooth and of Featured galaxies by selecting subjects with redshifts $1 \leq z \leq 3$, with surface brightnesses brighter than 24.5 mag~arcsec$^{-2}$, and which have ``star or artifact'' vote fractions (as described in Section \ref{sec:release}, from the {\small \tt t00\_smooth\_or\_featured\_a2\_artifact\_weighted\_frac} column) $f_{\rm artifact} < 0.4$. For the Smooth sample, we select galaxies having $f_{\rm smooth} > 0.6$; galaxies in the Featured sample have $\mffeat \geq 0.4$.

Each sample is further refined by matching to multi-wavelength $IJH$ bulge-disk decompositions of galaxies in the CANDELS fields by H\"au\ss ler et al. (in preparation) using the software package Galapagos-2, developed by the MegaMorph project \citep{haeussler13}. Each galaxy is fit with 2 \citet{sersic68} components: a bulge where the S\'ersic index $n$ is allowed to vary, and an exponential disk with fixed $n=1$. We choose galaxies where the bulge-disk fits reported no error flags and where the fit parameters converged to values well within the limits of constraints set by the fitting routine (i.e., limits on S\'ersic index and effective radius of $0.22 \leq n \leq 7.8$ and $0.33 \leq r_e \leq 390$, respectively). These selections result in samples of 51 Featured galaxies and 1,950 Smooth galaxies with reliable bulge-disk decompositions.

Additionally, we select local galaxies observed by SDSS at the same range of physical resolutions ($1.16 \leq {\rm FWHM} \leq 1.20$~kpc, corresponding to $0.0418 \leq z \leq 0.0436$ at median seeing in SDSS DR7) and classified as ``smooth'' and ``featured'' in Galaxy Zoo 2 according to the ``clean'' criteria described in that work \citep{willett13}. Within these constraints, we consider only galaxies with bulge-disk parametric fits by \citet{simard11}; the fits use an exponential disk and a free-$n$ bulge, as with the fits to CANDELS galaxies by H\"au\ss ler et al. This selects samples of 1169 Featured galaxies and 330 Smooth galaxies at $z \approx 0.04$. 

Figure \ref{fig:dM} shows the luminosities of these samples relative to the evolving $M^*$ of the luminosity functions measured at these redshifts. For the CANDELS samples we use the rest-frame $V$-band luminosity functions of \citet{marchesini12}, linearly interpolating between measured values of $M^*_V$ within $1 \leq z \leq 3$ to calculate $\Delta M_V \equiv M_{V} - M^*_V$ using rest-frame $V$-band absolute magnitudes for each galaxy. We compute a similar value for the SDSS samples, $\Delta M_g \equiv M_g - M^*_g$, using the local $g$-band luminosity function of \citet{blanton01} and rest-frame $g$-band absolute magnitudes of each galaxy. 

In order to meaningfully compare disk fractions between samples of different morphologies and redshifts, we require that both featured and smooth samples be matched in terms of how they sample the luminosity function of galaxies within their redshift ranges. Figure \ref{fig:dM} shows that the samples all have very different distributions of $\Delta M$. We therefore weight galaxies in each sample such that the weighted histograms of $\Delta M$ are statistically indistinguishable, as shown in Figure \ref{fig:dM_wt}. This method has the advantage of considering all the galaxies in the sample (with appropriate weights) rather than relying on randomly selected sub-samples.

We note that while Figure \ref{fig:dM} plots samples measured from within the same redshift ranges and data sets (CANDELS and SDSS, respectively) in the same panels, Figure \ref{fig:dM_wt} groups samples by Galaxy Zoo morphology. The weighted histograms in each panel are normalised with respect to each other so that no galaxy has a weight greater than 1; the samples are divided in this way because the goal is to compare smooth galaxies at $z \sim 0$ with smooth galaxies at $1 \leq z \leq 3$, and likewise for featured galaxies. Given these weightings, there are effectively 132 galaxies in each of the smooth samples, and 44 galaxies in each of the featured samples.

Figure \ref{fig:btot} shows the distribution of bulge-to-total luminosity ratios in the smooth (left panel) and featured (right panel) samples for CANDELS and SDSS. We use SDSS $g$-band results and interpolate the CANDELS multi-wavelength bulge-to-total ratios to the $g$ band (although we note that our qualitative results do not change if we use bulge-to-total ratios which are flux-summed to obtain one measurement across all observed bands). As expected, featured galaxies are generally disk-dominated. The differences in the bulge-to-total distributions between featured samples at $z \sim 0$ and $1 \leq z \leq 3$ is not significantly different: a Kolmogorov-Smirnov (K-S) test indicates a significance of $0.8 \sigma$ ($p_{KS} = 0.41$). We cannot rule out the hypothesis that these two distributions are drawn from the same parent population, despite the very different epochs they probe. 

However, the smooth galaxies at $1 \leq z \leq 3$ have a considerably more uniform distribution of bulge-to-total ratios than smooth galaxies at $z \sim 0$. While there is a population of disk-dominated smooth galaxies at $z \sim 0$, a significantly higher portion of the higher-redshift smooth galaxies have disk-dominated morphologies than in the local smooth population: 54 per cent of smooth galaxies at $1 \leq z \leq 3$ have B/Tot $<0.5$, versus 29 per cent of galaxies at $z \sim 0$. A K-S test confirms the differences in distributions are statistically significant, indicating the distributions are inconsistent with being drawn from the same parent population at the $4.3 \sigma$ level ($p = 1.5 \times 10^{-5}$).

Smooth disk-dominated galaxies are typically more compact than featured disk-dominated galaxies in the CANDELS samples: the median 80 percent flux radius, $r_{80}$, of the weighted smooth sample with disk-dominated bulge-to-total ratios is 5~kpc, compared with 7.5~kpc for the featured disk-dominated sample. Within the quoted significant figures, the median $r_{80}$ values are not sensitive to the choice of B/Tot used to define ``disk-dominated''. Disk-dominated galaxies with smooth morphologies are more compact, on average, than disk-dominated galaxies with clear features, at the same luminosities and colours.

This finding of a substantial population of completely smooth disk galaxies at $z > 1$ is consistent with a previous study of a much smaller sample \citep{conselice11}, with the observed decline in disk bar fraction at $z \gtrsim 1$ \citep{melvin14,simmons14}, and also with the results of recent dynamical studies of galaxies at $z > 1$ \citep[e.g.][]{wisnioski15} showing that disk galaxies are on average dynamically warmer and thus less prone to instabilities such as bar and spiral modes. In the absence of such features these disks are likely to be classified as ``smooth''.

The existence of this population makes the common task of selecting a sample to study the properties of disk galaxies more difficult. Figure \ref{fig:btot_colour} shows the distributions of B/Tot for smooth galaxies split according to colour-based selection criteria. A selection on $U-V$ colour is ineffective at separating disk-dominated from bulge-dominated smooth galaxies. Additionally incorporating $V-J$ colours so as to sort smooth galaxies into quiescent and star-forming categories \citep[drawn as dashed lines Figure \ref{fig:UVJ}, according to empirical criteria from][]{williams09} is somewhat more effective, but is still far from ideal. If one were to select star-forming smooth galaxies as a proxy for selecting disk-dominated smooth galaxies from the weighted samples shown in Figures \ref{fig:dM_wt} and \ref{fig:btot}, the selection would include 77 per cent of smooth galaxies with B/Tot $<0.5$, but would also include a substantial fraction of smooth galaxies with B/Tot $>0.5$, such that the sample would only be 63 per cent pure. If, on the other hand, one were to select quiescent smooth galaxies as a proxy for spheroid-dominated smooth galaxies, that selection would only be 48 per cent complete at identifying galaxies with B/Tot $>0.5$ and 64 per cent pure. The issue is not unique to high-redshift galaxies: using colour as a proxy for light distribution in smooth galaxies from the SDSS sample described above \citep[and using K-corrected colours from][]{blanton05} leads to similarly low completeness and purity fractions for lower-redshift smooth-disk samples. Colour-based criteria alone are not particularly efficient at separating disk-dominated from spheroid-dominated galaxies that have mostly smooth morphologies. 

Disk galaxies with smooth light distributions will be most cleanly selected by using a cut on the bulge-to-total ratio. However, those with complex features in the disk, whether these are spiral arms at low-redshift or clumpy galaxies in higher-redshift samples, may suffer from catastrophic failure of fitting if modelled with smooth distributions. Therefore, a superset of galaxies selected from both Galaxy Zoo morphologies and smooth disks from cuts on bulge-to-total ratios will provide a more complete sample of $z > 1$ disks than either method alone. A complementary selection may be used to identify a cleaner sample of disk-free smooth galaxies. The subset of 1,149 smooth galaxies at $1 \leq z \leq 3$ with high-quality fitting and B/Tot $<0.5$ is flagged as an extra column in Table \ref{table:data-main}, available as part of this Galaxy Zoo CANDELS data release.

%%%%%%%%%%%%%%%%%%%%%%%%%%%%%%%%%%%%%%%%%%%%%%
%
%  
\section{Summary}\label{sec:summary}
%
%
%%%%%%%%%%%%%%%%%%%%%%%%%%%%%%%%%%%%%%%%%%%%%%

The Galaxy Zoo project has collected typically 40 or more independent visual classifications to date from colour images of three CANDELS fields: GOODS-South, COSMOS, and the UDS. Here we present the public release of these classifications, both in raw form and after applying an iterative consensus-based classifier weighting scheme that has been successfully applied to multiple previous Galaxy Zoo projects, as well as additional weighting techniques making use of the stellarity parameter from automated measurements. We provide an analysis of changes in classifications with imaging depth and offer advice and caveats for usage of these morphological measurements for different science goals. 

Approximately 12 per cent of the full catalog is comprised of galaxies where at least 30 per cent of weighted classifiers indicated the galaxy was ``featured''. These galaxies include clumpy galaxies, edge-on galaxies, galaxies with spiral arms and with bulges at a range of strengths. Each of these morphological parameters is reported for every galaxy in the sample. 

Comparison of the Galaxy Zoo morphologies with existing visual morphologies from other studies shows remarkably good agreement across a wide range of morphological features. We also combine Galaxy Zoo morphologies with multi-wavelength bulge-disk decompositions to show that, while the presence of features is a reliable indicator of a disk, the \emph{absence} of such features does not imply an absence of a significant disk. A substantial fraction of galaxies lacking significant morphological signatures of disk features have disk-dominated light profiles; this fraction is significantly higher at $z > 1$ than for luminosity- and resolution-matched samples of smooth galaxies at $z \sim 0$. We identify a sample of smooth disks at $1 \leq z \leq 3$ as part of this Galaxy Zoo CANDELS data release.

This data release includes 3 of the 5 legacy fields imaged by the CANDELS project. Future work in the collaboration between Galaxy Zoo and CANDELS includes the classification of the other 2 fields (EGS and GOODS-North). Other future studies of this rich dataset may include a more detailed study of clumpy structures in $z > 1$ galaxies and examination of mergers at these epochs, for comparison to local samples with similar selection methods \citep{darg10b,darg10a}. 

The public catalog of Galaxy Zoo CANDELS morphologies may be obtained from {\tt data.galaxyzoo.org}.
  
%%%%%%%%%%%%%%%%%%%%%%%%%%%%%%%%%%%%%%%%%%%%%%
%%%%%%%%%%%%%%%%%%%%%%%%%%%%%%%%%%%%%%%%%%%%%%
%%%%%%%%%%%%%%%%%%%%%%%%%%%%%%%%%%%%%%%%%%%%%%
%%%%%%%%%%%%%%%%%%%%%%%%%%%%%%%%%%%%%%%%%%%%%%
%
%
\section*{Acknowledgments}
%
%
%%%%%%%%%%%%%%%%%%%%%%%%%%%%%%%%%%%%%%%%%%%%%%

The authors wish to thank the anonymous referee for helpful comments which improved this manuscript. Astropy \citep{astropy13}, TOPCAT \citep{taylor05} and an OS X widget form of the JavaScript Cosmology Calculator \citep{wright06,rsimpson13} were used while preparing this paper. Figure \ref{fig:sankey} was created using {\tt SankeyMATIC} at {\tt sankeymatic.com/build}. 
BDS gratefully acknowledges support from the Oxford Martin School and Balliol College, Oxford. Support for this work was provided by the National Aeronautics and Space Administration through Einstein Postdoctoral Fellowship Award Number PF5-160143 issued by the Chandra X-ray Observatory Center, which is operated by the Smithsonian Astrophysical Observatory for and on behalf of the National Aeronautics Space Administration under contract NAS8-03060.
SJK acknowledges funding from the STFC Grant Code ST/MJ0371X/1.
We thank M. Schwamb and the ASIAA for hosting the ``Citizen Science in Astronomy'' workshop, 3-7 Mar 2014 in Taipei, Taiwan, at which some of this analysis was done.

The development of Galaxy Zoo was supported in part by the Alfred P. Sloan Foundation. Galaxy Zoo was supported by The Leverhulme Trust. 

This work is based on observations taken by the CANDELS Multi-Cycle Treasury Program with the NASA/ESA HST, which is operated by the Association of Universities for Research in Astronomy, Inc., under NASA contract NAS5-26555.
  
\bibliographystyle{mn2e}
\bibliography{refs}

\end{document}